\documentclass[a4paper,fleqn,usenatbib]{mnras}

\usepackage{newtxtext,newtxmath}

\usepackage[T1]{fontenc}
\usepackage{ae,aecompl}

\usepackage[nolist,nohyperlinks,printonlyused]{acronym}

\usepackage{float}

\usepackage{graphicx}	
\usepackage{amsmath}	
\usepackage{comment}

\newcommand{\HI}{\ion{H}{I}}

\newcommand{\mathHI}{{\mbox{\scriptsize \HI}}}

\newcommand\given[1][]{\:#1\vert\:}

\newcommand{\lyaf}{\text{Ly$\alpha$ forest}}
\newcommand{\lya}{\text{Ly$\alpha$}}

\newcommand{\NHI}{$N_{\mathHI{}}$}

\newcommand{\vpfit}{\texttt{VPFIT}}
\newcommand{\bndist}{$b$-$N_{\mathHI{}}$~distribution}
\newcommand{\bn}{$\left\{b, N_{\mathHI{}}\right\}$}
\newcommand{\dndz}{d$N$/d$z$}

\defcitealias{Danforth2016}{D16}
\defcitealias{Hu2022}{Hu22}

\title[Impact of the WHIM on IGM Thermal State]{The Impact of the WHIM on the IGM Thermal State Determined from the Low-$z$ Lyman-$\alpha$ Forest}

\author[Hu et al.]{
Teng Hu,$^{1}$\thanks{E-mail: tenghu@ucsb.edu (UCSB)}
Vikram Khaire$^{1,2}$,
Joseph F. Hennawi$^{1,3}$,
Jose O\~norbe$^{4}$,
Michael Walther$^{5}$,\newauthor
Zarija Lukic$^{6}$ and
Frederick Davies$^{7}$
\\
$^{1}$Physics Department, Broida Hall, University of California Santa Barbara, Santa Barbara, CA
93106-9530, USA\\
$^{2}$ Indian Institute of Space Science \& Technology, Thiruvananthapuram, Kerala - 695547, INDIA\\
$^{3}$Leiden Observatory, Leiden University, PO Box 9513, NL-2300 RA Leiden, the Netherlands\\
$^{4}$Facultad de F\'isica, Universidad de Sevilla, Avda. Reina Mercedes s/n, Campus de Reina Mercedes, E-41012 Sevilla, Spain\\
$^{5}$University Observatory, Faculty of Physics, Ludwig-Maximilians-Universität München, Scheinerstr. 1, 81679 Munich, Germany\\
$^{6}$Lawrence Berkeley National Laboratory, Berkeley, CA 94720, USA\\
$^{7}$Max-Planck-Institut für Astronomie, Königstuhl 17,69117 Heidelberg, Germany\\
}

\date{Accepted XXX. Received YYY; in original form ZZZ}

\pubyear{2015}

\begin{document}
\label{firstpage}
\pagerange{\pageref{firstpage}--\pageref{lastpage}}
\maketitle

\begin{abstract}

At $z \lesssim 1$, shock heating caused by large-scale velocity flows and possibly violent feedback from galaxy formation, converts a significant fraction of the cool gas ($T\sim 10^4$ K) in the intergalactic medium (IGM) into warm-hot phase (WHIM)  with $T >10^5$K, resulting in a significant deviation from the previously tight power-law IGM temperature-density relationship, $T=T_0 (\rho\slash {\bar{\rho}})^{\gamma -1}$. 
This study explores the impact of the WHIM on measurements of the low-$z$ IGM thermal state, $[T_0,\gamma]$, based on the \bndist{} of the \lyaf{}.  
Exploiting a machine learning-enabled simulation-based inference method trained on Nyx hydrodynamical simulations, we demonstrate that [$T_0$, $\gamma$] can still be reliably measured from the \bndist{} at $z=0.1$, notwithstanding the substantial WHIM in the IGM. 
To investigate the effects of different feedback, we apply this inference methodology to mock spectra derived from the IllustrisTNG and Illustris simulations at $z=0.1$. 
The results suggest that the underlying $[T_0,\gamma]$ of both simulations can be recovered with biases as low as $|\Delta \log(T_0/\text{K})| \lesssim 0.05$ dex,  $|\Delta \gamma | \lesssim 0.1$, smaller than the precision of a typical measurement. 
Given the large differences in the volume-weighted WHIM fractions between the three simulations (Illustris 38\%, IllustrisTNG 10\%, Nyx 4\%) we conclude that the \bndist{} is not sensitive to the WHIM under realistic conditions. 
Finally, we investigate the physical properties of the detectable \lya{} absorbers, and discover that although their $T$ and $\Delta$ distributions remain mostly unaffected by feedback, they are correlated with the photoionization rate used in the simulation.
\end{abstract}




\section{Introduction}
  \label{sec:intro}

Being the largest reservoir of baryons in the Universe, 
the \ac{IGM} plays a crucial role in the evolution of the Universe and the formation of structures. 
Based on the canonical cosmological model constrained by many
observational studies, the thermal evolution of the \ac{IGM} is dominated by 
two major phase transition events of the Universe.
The first phase transition is hydrogen reionization caused by the first generation of galaxies at redshift $6 < z < 20$ \citep{Becker2001, Fan2006, Robertson2015, mcgreer1}. 
The second one is the double reionization of Helium
(\ion{He}{II}$\rightarrow$\ion{He}{III}) driven by \ac{QSO}s \citep[see e.g.][]{MadauMeiksin1994, mcquinn09, Khaire2017}, 
which is believed to occur at $z \sim 3$ \citep[][]{Worseck2011, Syphers2014,  Worseck2018}, 
where the quasar luminosity function reaches its peak
\citep[see e.g.][]{Hopkins2007, Khaire15puc, Kulkarni2019}.
These two-phase transition events heat up the \ac{IGM} dramatically to a maximum of 15,000K while ionizing the \ac{IGM}.

After the completion of hydrogen reionization ($\Delta z \sim $1-2), 
the \ac{IGM} thermal state is shaped by the quasi-equilibrium balance between photoionization heating from the extragalactic UV background \citep{H&M2012, KS19}
and various cooling processes including recombinations, excitation, cooling due to Hubble expansion, 
and inverse Compton scattering of electrons off of the cosmic microwave background \citep[CMB; see e.g.][]{McQuinn2016review}.  
All these processes together drive the \ac{IGM} to follow a power-law temperature-density ($T$-$\Delta$) relation: 
\begin{equation}
T(\Delta) = T_0 \Delta^{ \gamma -1},
\label{eqn:rho_T}
\end{equation}
where $\Delta = \rho/ \bar{\rho}$ is the overdensity,
$T_0$ is the temperature at mean density, and $\gamma$ is the adiabatic index \citep{hui1, McQuinn2016}.
These two parameters $[T_0,\gamma]$ thus characterize the thermal state of the \ac{IGM}, 
making it feasible to
constrain the \ac{IGM} thermal history \citep{Lidz2010, Becker2011, Rorai2017, Hiss2018, WaltherM2019, Gaikwad2021} by measuring [$T_0$, $\gamma$] at different epochs.
These measurements improve our knowledge of the \ac{IGM} thermal evolution and shed light on the underlying heating and cooling processes involved.

Nevertheless, the aforementioned power-law $T$-$\Delta$ relationship for the \ac{IGM} could potentially break down at $z \lesssim 1$, where shock heating caused by large-scale velocity flows \citep{Cen2006,Nath2001} 
and various feedback mechanisms become more common \citep{Scannapieco2005,Khaire2023}.
Specifically, shock heating at low-$z$ converts a notable fraction of the cool IGM into \ac{WHIM} with $T >10^5$K \citep{Shull2012}, causing a substantial dispersion in the IGM $T$-$\Delta$ distribution \citep{Dave2001, Cen2006}. 
As a result of such dispersion,
the IGM $T$-$\Delta$ distribution can no longer be fully described by the typical power-law relationship (see Fig.~\ref{fig:tri_rho_T}), which introduces additional complexities in the measurement of the IGM thermal state \citep[][hereafter \citetalias{Hu2022}]{Hu2022}.
The imperative question is whether the significant shock heating at low-$z$ influences the observable, i.e., the \lyaf{}, which serves as the primary probe of the IGM, and if it does, how might such impacts affect measurements of the IGM thermal state $[T_0,\gamma]$?

In practice, the IGM thermal state can be measured through various statistical properties of the \lyaf{}. 
Particularly, at $z\lesssim 3$, the \lyaf{} is amenable to Voigt profile decomposition \citep[see][]{Hiss2018}, where each line can be fit by three parameters: redshift $z_{\text{abs}}$, Doppler broadening $b$, and neutral hydrogen column density $N_{\rm HI}$. 
The IGM thermal state at these redshifts can thus be measured using the 2D joint \bndist{}
\citep{schaye1999,schaye2000,bolton2014, Rorai2018,Hiss2018}.
\citetalias{Hu2022} introduced a new inference method to measure the thermal state [$T_0$, $\gamma$] and the photoionization rate $\Gamma_{\mathHI{}}$ of the IGM based on the \bndist{} and \lya{} line density, \dndz{}, of the \lyaf{}.
Such a method performs Bayesian inference with the help of neural networks and Gaussian emulators, trained on a suite of Nyx simulations \citep{Almgren2013,Lukic2015}, making it possible to measure the thermal state of the IGM to high precision for realistic mock datasets.

Moreover, the thermal state of the IGM at $z <1.7$ remains poorly constrained, since the \lya{} transition below such redshift lies below the atmospheric cutoff ($\lambda \sim 3300 \text{\AA{}}$), requiring UV observations from space with \ac{HST}.
After He II reionization ($z <3$), the thermal state of the IGM is considered to be dominated by adiabatic cooling from Hubble expansion, which leads to an \ac{IGM} thermal state with $ T_0\sim 5000$K and $\gamma \sim 1.6$ at the current epoch $z=0$ \citep{McQuinn2016}. 
However, such a prediction of low temperatures has not yet been confirmed observationally. 
Meanwhile, recent studies have suggested that the \lya{} lines appear broader than predicted by numerical simulations at $z <0.5$ \citep{Gaikwad2017, Viel2017, Nasir2017}. This observation is based on the $b$ parameters acquired from the \ac{HST} \ac{COS} spectra \citep[][referred hereafter as \citetalias{Danforth2016}]{Danforth2016} dataset. 
While it has been argued that such a mismatch might be resolved by additional sources of turbulence,
an alternative explanation would be that the \ac{IGM} is actually hotter than previously presumed, with $T_0$ conceivably approaching $10,000$K, 
implying the existence of unexpected sources of heating \citep{Bolton2021, Bolton22},
which, if true, would change our understanding of the \ac{IGM} physics thoroughly.


In this paper, we adopt the \citetalias{Hu2022} inference method to investigate the impact of the WHIM on measurements of the IGM thermal state, $[T_0,\gamma]$, based on the \bndist{} of the \lyaf{}.
Firstly, we assess the effectiveness of $[T_0,\gamma]$ as IGM parameters at low-$z$ by comparing its performance as neural network training labels against the photoheating labels [$A$,$B$] (see \S~\ref{sec:Grid_label}). 
These latter labels are photoheating rate rescaling factors used to generate the Nyx simulation suite with various thermal histories \citep[see e.g][]{Becker2011}. 
Since our emulators are trained on these Nyx simulations generated by varying $[A,B]$, the inference method is naturally inclined to retrieve these photoheating labels. 
On the other hand, if shock heating at low-$z$ causes the $T$-$\Delta$ distribution of the \lya{} absorbers to deviate from the power-law relationship, the effectiveness of $[T_0,\gamma]$ as labels could be compromised.
Thus, our comparison between these two sets of labels provides insight into the robustness of $[T_0,\gamma]$ as IGM parameters at low-$z$, in the presence of substantial shock heating.

Afterwards, we explore the potential effects of different feedback mechanisms, which are associated with WHIM, on measurements of the IGM thermal state, $[T_0,\gamma]$.
In terms of our inference methodology, the question becomes: what would happen if we used a simulation grid without feedback to interpret a Universe that includes feedback? Would this lead to unbiased $[T_0,\gamma]$?
To answer these questions, we apply the \citetalias{Hu2022} inference methodology to mock data drawn from the Illustris \citep{Illustris} and IllustrisTNG \citep{IllustrisTNG} simulations at $z=0.1$. These two simulations incorporate galaxy formation models and feedback mechanisms that are not included in the Nyx simulation, which heat up the IGM substantially at low-$z$, and transform the cool diffuse \lya{} gas into WHIM more effectively compared with Nyx simulation (see Fig.~\ref{fig:tri_rho_T}). We examine the inference results based on these two simulations and explore whether feedback biases the measurement of the thermal state [$T_0$, $\gamma$].

To further investigate this problem, we explore the specific impacts of shock heating and other astrophysical processes, such as AGN feedback and UV background photoionization, on the physical properties of the \lyaf{} at $z = 0.1$. Within the three aforementioned simulations, we identify simulated \lya{} absorbers in the simulations and establish a direct correlation between the physical properties of these absorbers (including temperature $T$, overdensity $\Delta$, and line-of-sight velocity $v_\text{los}$) and the observed \lya{} line parameters ($b$, $N_{\mathHI{}}$) derived from the absorption lines detected in corresponding mock spectra.
We then examine the distributions of $\Delta$ and $T$ of these simulated \lya{} absorbers across the three aforementioned simulations to study the detailed effects of the feedback and UV background photoionization rate, $\Gamma_{\mathHI{}}$, on the \lyaf{}.

This paper is organized as follows:
In section \S~\ref{sec:simulations}, we outline the simulations and associated processes
applied to generate synthetic \lya{} forest.  It includes post-processing, forward-modeling, and Voigt profile fitting. 
The inference framework and results for all three simulations are then presented in
in \S~\ref{sec:inference}.
Section \S~\ref{sec:lowz_lyaf} is dedicated to the investigation of the physical characteristics of low-redshift \lyaf{} absorbers in all three simulations.
Finally, in \S~\ref{sec:discussion}, we present a summary and discussion of our findings. For the sake of brevity, we use $\log$ as a shorthand to denote $\log_{10}$ throughout the paper.

\section{Simulations}
  \label{sec:simulations}
In this paper, we utilize the inference framework described in \citetalias{Hu2022},
which employs the \bndist{} emulator built on neural networks trained on a set of Nyx simulations.
We also use galaxy formation simulations IllustrisTNG and Illustris to investigate the low-$z$ \lyaf{} under different feedback mechanisms.
Since this work focuses on the low redshift \lyaf{}, we use $z=0.1$ simulation snapshots for all three simulations. 
In this section, we first provide a description of the simulations and the implemented physical models, followed by the (mock) data processing procedures employed in our study. This includes the generation of simulated line-of-sight (LOS) of \lya{} forest (hereafter referred to as skewers for simplicity),
forward modelling, and the Voigt profile fitting of \lya{} lines.
The cosmological parameters and thermal states of the three simulations are summarized in Table~\ref{tab.cosmo}. 

\begin{table}
\centering
\caption{Parameters of cosmology and $T-\Delta$ relation (at $z=0.1$)}
\begin{tabular}{cccc}
\hline
Parameters  &  Nyx   & IllustrisTNG & Illustris  \\ 
\hline
${\Omega_m}$     & 0.3192  &  0.3089  &  0.2726\\
$\Omega_{\Lambda}$&0.6808 &  0.6911   &  0.7274\\
$\Omega_{b}$     & 0.0496  &  0.0486  &  0.0456 \\ 
$h$              &0.670   &  0.6774  &  0.704  \\
$\sigma_{8}$     &0.8288   &  0.8159  &  0.809  \\
$n_s$            &0.96    &   0.97   &  0.963\\
\hline
$T_0$            &4093 K  &  4241 K  &  4292 K  \\
$\gamma$         &1.588  &  1.593   &    1.577  \\
 \hline
\end{tabular}
\label{tab.cosmo}
\end{table}
\subsection{Nyx}
\label{sec:Nyx}

Nyx is an adaptive mesh, massively parallel, cosmological simulation code primarily developed to simulate the \ac{IGM} \citep{Almgren2013,Lukic2015}.
Nyx simulates the dark matter evolution by treating the dark matter 
as self-gravitating Lagrangian particles,
while it models baryons as an ideal gas on a 
uniform Cartesian grid following an Eulerian approach.
The Eulerian hydrodynamics equations are solved using a second-order piece-wise parabolic method,
which is capable of accurately capturing shocks. 

Nyx includes the major astrophysical processes relevant to the evolution of the 
the \lyaf.
First, gas in the Nyx simulation is treated as having a primordial composition with a hydrogen mass fraction of 0.76 and helium mass fraction of 0.24,
and zero metallicity. 
Nyx takes into account the process of inverse Compton cooling off the microwave background and tracks the total thermal energy loss due to atomic collisional processes.
Nyx also implements recombination, collisional ionization, dielectric recombination, and cooling following the prescription given in \citet{Lukic2015}. 
Ionizing radiation in Nyx is modelled by a spatially uniform but time-varying ultraviolet background radiation field of \citet[]{H&M2012},
while assuming all cells in the simulation are optically thin.
Furthermore, following standard practice, we allow the UV background photoionization rate, $\Gamma_{\mathHI{}}$, 
to be a free parameter in post-processing while generating mock \lya{} skewers. 
Lastly, Nyx does not implement any galaxy formation or feedback,
which simplification reduces the required computational resources  significantly, 
allowing us to run a large ensemble of simulations with different thermal histories (see \S~\ref{sec:Grid_label}), which is required for accurate statistical inference. 

Each Nyx simulation model used in this study was
initialized with the same initial condition at $z=159$ and evolved down to $z=0.03$ in a $L_{\text{box}} = 20~{\rm cMpc}\slash h$ simulation box with $N_{\text{cell}} = 1024^3$ Eulerian cells and $1024^3$ dark matter particles. 
The box size is a compromise between computational cost and the need for convergence at least to $< 10\%$ on small scales (large $k$). 
In short, such choices of box size and resolution should not affect the line parameters of the \lyaf{} significantly.
More discussion of the resolution, box size, and convergence issues can be found in \citet{Lukic2015} and \citet{Hu2022}.

  \subsection{IllustrisTNG and Illustris}
\label{sec:ILL}
\begin{figure*}
 \centering
    \includegraphics[width=1.00\linewidth]{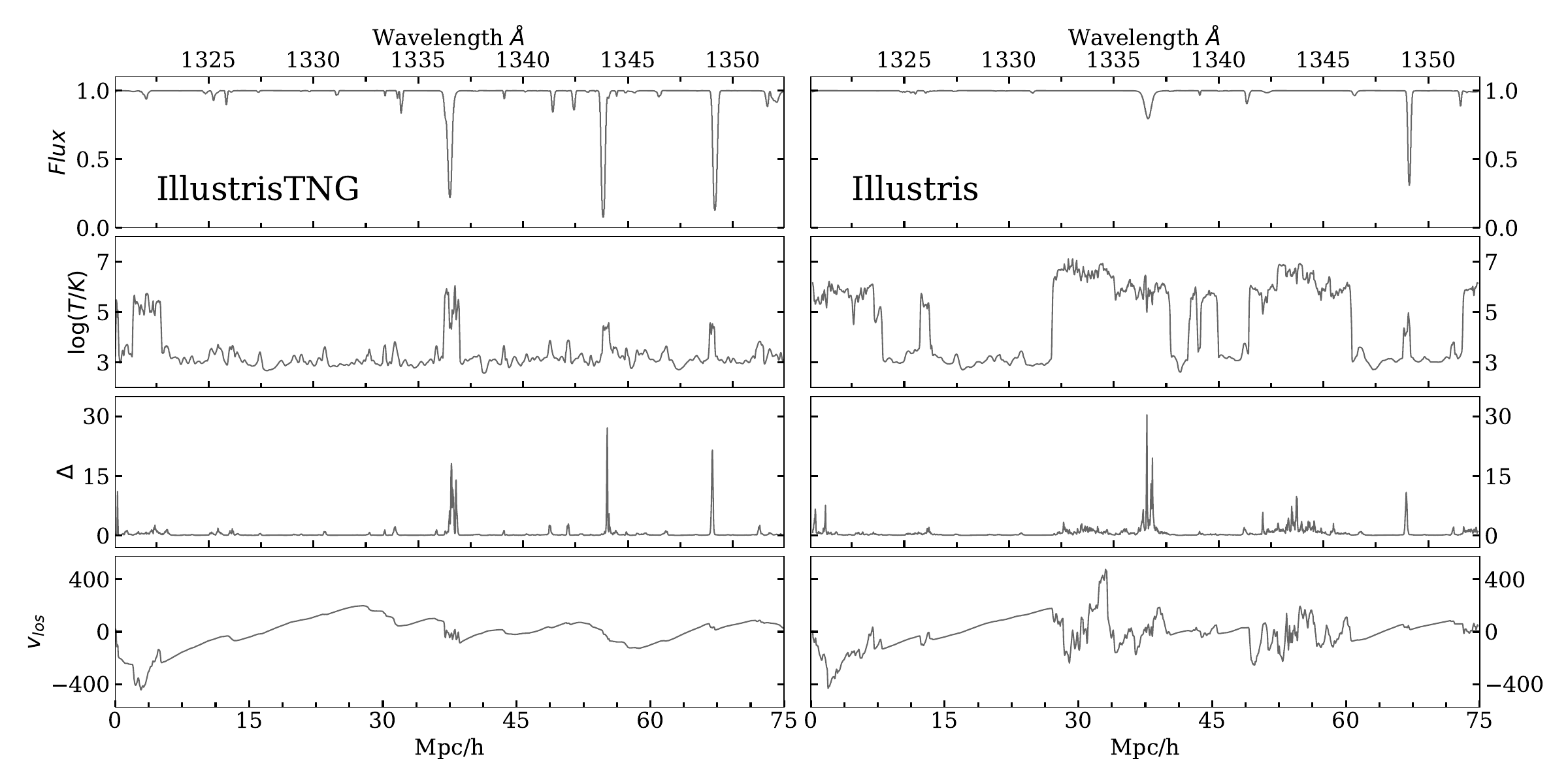}
  \caption{Examples of simulation skewers for IllustrisTNG (left) and Illustris (right) simulations, probing the structure generated by the same initial condition, while the two simulations are post-processed to share the same UV backgrounds photoionization rate, $\Gamma_{\mathHI{}}$. The flux is plotted in black on the top panel, while the temperature $T$, overdensity $\Delta$, and velocity along LOS $v_\text{los}$ are shown in black in the second, third, and bottom panels consecutively.
  }
  \label{fig:ILL_skewer}
\end{figure*} 

To evaluate the effectiveness of the IGM thermal state [$T_0$, $\gamma$] as the IGM parameters and test the efficacy of our inference framework on the realistic IGM, which can be affected by astrophysical processes that are not included in Nyx simulation such as galaxy formation and AGN feedback,
we employ Illustris \citep{Illustris} and IllustrisTNG \citep{IllustrisTNG} simulations, and use them as mock observational data in our inference method.

The IllustrisTNG and Illustris are cosmological hydrodynamic simulations powered by the {\sc arepo} code \citep{Springel2010}. This code employs a moving mesh approach to solve hydrodynamics through the Euler equations, and it computes gravitational forces on a quasi-Lagrangian moving Voronoi mesh via the tree-PM method. Both simulations incorporate a wide range of astrophysical processes for galaxy formation, such as star formation, stellar and AGN feedback, galactic winds, and chemical enrichment. They utilize the UV background detailed in \cite{Faucher-Gigu2009} for photoionization heating and cooling. Other processes for modelling the \lya~forest, like collisional ionization and inverse Compton cooling from the cosmic microwave background, are also taken into account.

The primary distinction between IllustrisTNG and Illustris lies in their AGN feedback mechanisms, especially regarding AGN feedback. Both simulations implement AGN feedback in two modes based on the gas accretion rate onto the central supermassive black hole: the `quasar-mode' at high accretion rates \citep{Springel05, Hopkins08, Debuhr11} and the `radio-mode' at low rates \citep{Croton06, Bower06, Sijacki07}. While both use continuous thermal feedback in `quasar-mode', their `radio-mode' implementations differ. Illustris employs a bubble model for radio-mode feedback, accumulating substantial feedback energy for explosive release, often ejecting excessive hot gas \citep{Illustris}. Conversely, IllustrisTNG models this feedback as a kinetic wind, injecting momentum into neighbouring regions from the central black hole. This approach better replicates astrophysical properties like star formation rates and galaxy colour distributions \citep{Dylan18_color, Pillepich18}.

Both the IllustrisTNG and Illustris simulations we used in this study have box sizes of $75$ cMpc/h and $1820^3$ baryon and dark matter particles. 
Since {\sc arepo} is a moving mesh code,
we convert the Voronoi mesh outputs to $1820^3$ cartesian grids by dumping the smoothed quantities such as temperature, density, and velocities on grids to generate 
\lya{} forest skewers. A Gaussian kernel with a size equal to 2.5 times the radius of each Voronoi cell is applied for the smoothing, assuming each Voronoi cell is spherical. We then generate skewers for IllustrisTNG and Illustris simulations following the approach discussed in \S~\ref{sec:FM_VPFIT}.
In Fig.~\ref{fig:ILL_skewer}, we plot two simulation skewers for IllustrisTNG and Illustris respectively, while the two simulations are post-processed to share the same UV backgrounds photoionization rate, $\Gamma_{\mathHI{}}$ (see \S~\ref{sec:GUV_dNdz} for more discussion).
The flux ($e^{-\tau}$) is plotted in the top panel, and the temperature, over-density, and line-of-sight velocity profiles are shown in the second, third, and bottom panels consecutively.
It is worth mentioning that the two skewers probe the structure generated by the same initial condition, and the difference in $T$, $\Delta$, and $v_\text{los}$ are caused by different feedback strengths, i.e., the Illustris exhibits higher temperatures due to its stronger feedback, which results in weaker absorption features given the same UV backgrounds.
More discussion on the difference between \lya{} forest in IllustrisTNG and Illustris simulations can be found in \citet{Khaire2023}.

\subsection{IGM Thermal State and Parameter Grid}
\label{sec:Grid_label}

\begin{figure*}
 \centering
    \includegraphics[width=1.00\linewidth]{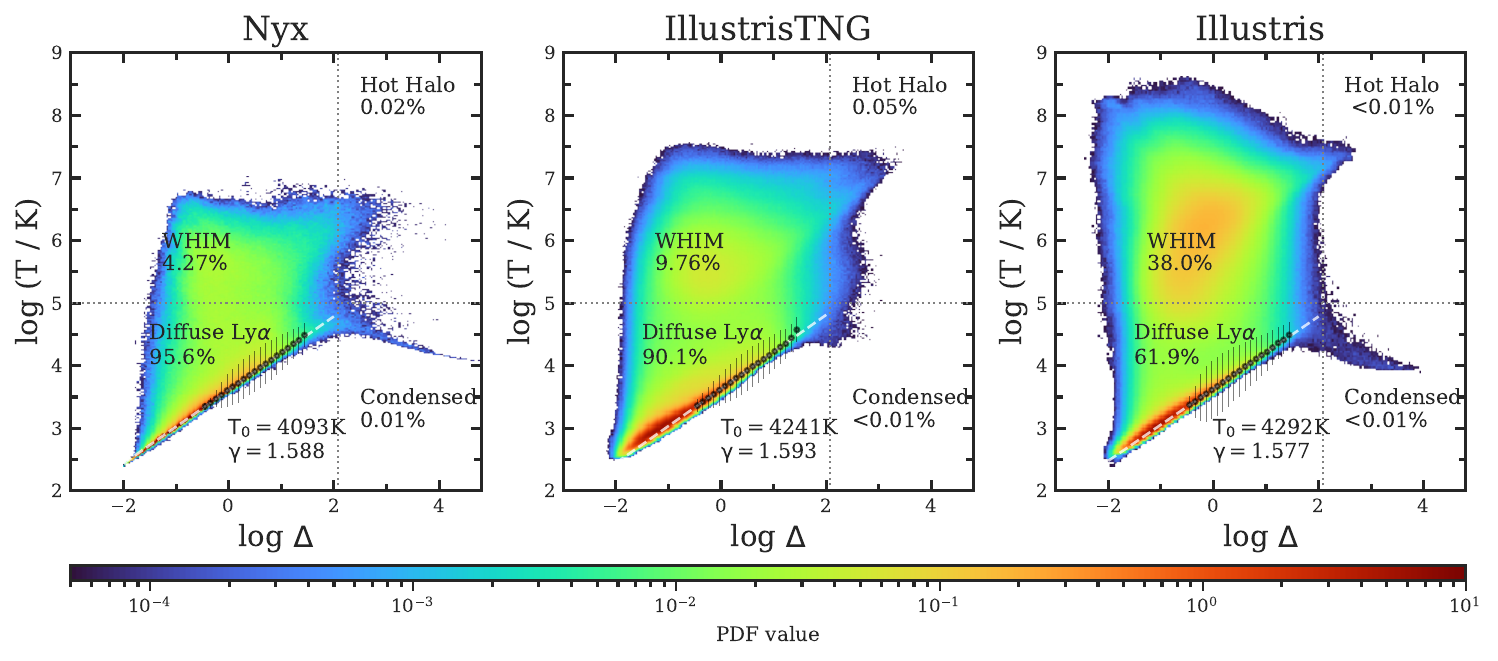}
  \caption{ Volume weighted $T$-$\Delta$ distribution for all three simulations at $z=0.1$. 
The $\log T$ for each bin are plotted as black dots, and the 1-$\sigma_{T}$ error bars are shown as black bars. The best-fit power-law relationship is shown as white dashed lines.
The Nyx (left) model is the default model which has {$\log (T_0/\text{K})$} = 3.612, $\gamma=1.588$; and IllustrisTNG (middle) yields {$\log (T_0/\text{K})$} = 3.627, $\gamma=1.593$; whereas Illustris (right)  has {$\log (T_0/\text{K})$} = 3.633, $\gamma=1.577$. The gas phase fractions are shown in the annotation. }
  \label{fig:tri_rho_T}
\end{figure*}

 \begin{figure*}
 \centering
    \includegraphics[width=1.0\linewidth]{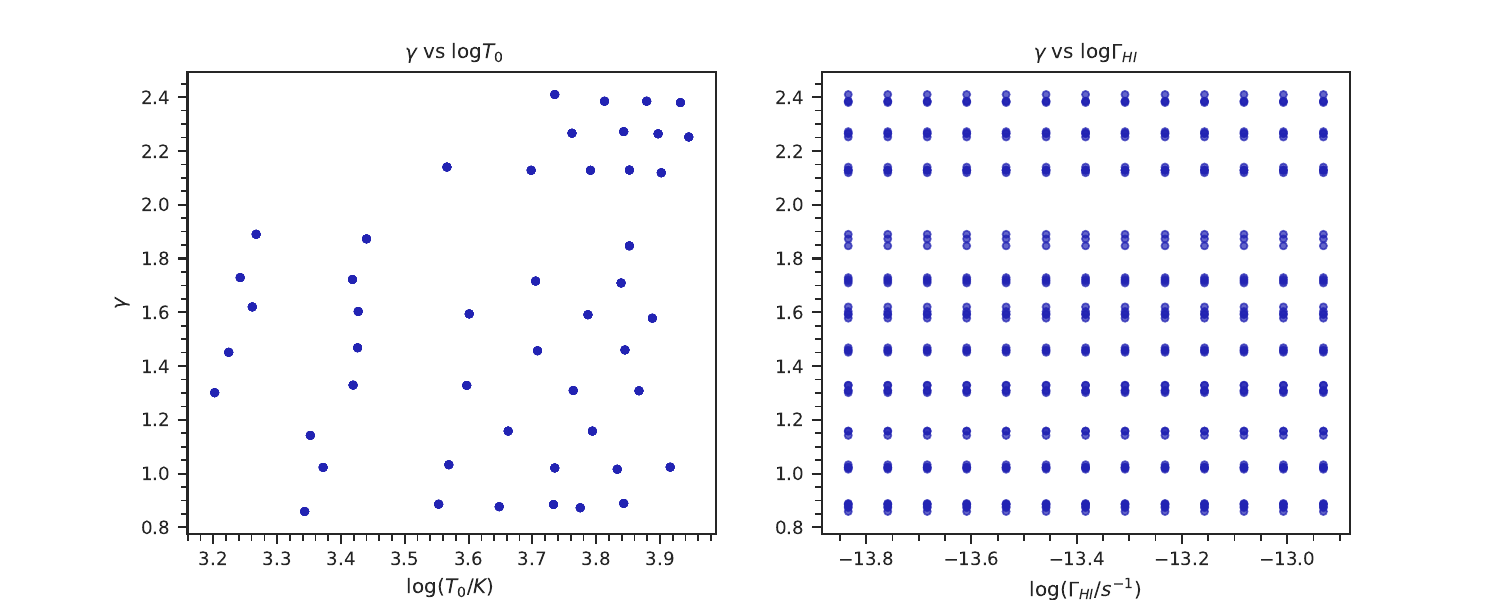}
  \caption{ Parameters grid (blue circles) from snapshots of Nyx simulations from the \ac{THERMAL} suite at $z=0.1$, parameterized by the thermal state [$T_0$,$\gamma$]. 
  The left panel shows the $\gamma$ - $T_0$ grid, 
  whose shape is determined by the photoheating labels [$A$,$B$] (see Fig.~\ref{fig:z01_AB_grid}) and
  the evolution of the thermal state of the IGM. 
  The right panel is $\gamma$ - $\Gamma_{\mathHI{}}$ grid, showing the 13 $\Gamma_{\mathHI{}}$ values for each point on the $\gamma$ - $T_0$ grid. }
  \label{fig:lowz_grid}
\end{figure*}

Following \citetalias{Hu2022}, we make use of the \ac{THERMAL}\footnote{For details of THERMAL suite, see http://thermal.joseonorbe.com} suite of Nyx simulations \citep[see][]{Hiss2018,WaltherM2019} to model the IGM with various thermal histories.
The suite consists of 48 models with varying thermal histories,
each generated by changing the photoheating rate of the simulation following the prescription described in \cite{Becker2011},
in which the photoionizing rate, $\epsilon$ is assumed as a function of overdensity, i.e. 
\begin{equation}
\epsilon =\epsilon_{\rm HM12}(z) A\Delta^B, 
\label{eqn:heating}
\end{equation}
where $\epsilon_{\rm HM12}(z)$ 
stands for the time-varying photoheating rate per H~{\sc ii} ion
tabulated in 
\cite{H&M2012}, and the constants $A$ and $B$ are free photoheating parameters that are varied in the different Nyx runs to achieve different thermal histories,
which results in different thermal states at $z=0.1$.
The distribution of parameters in our thermal grid, i.e. the different 
values of [$T_0$, $\gamma$] are illustrated in Fig~\ref{fig:lowz_grid}, and the corresponding values of [$A$, $B$] are presented in Fig.~\ref{fig:z01_AB_grid} (see Appendix~\ref{sec:label_AB} for more discussion).

Conventionally, the thermal parameters [$T_0$,$\gamma$] are obtained by fitting a power law to the $T$-$\Delta$ relationships (See eq.~\ref{eqn:rho_T}).
Such a fitting procedure is straightforward at higher redshift ( $z \gtrsim 2$) where the $T$-$\Delta$ distributions of the IGM are tight. However, in low-$z$, the distributions of the IGM temperatures are noticeably broader due to the extensive shock heating, which heats up the IGM, resulting in more \ac{WHIM}. 
The $T$-$\Delta$ distributions for all three simulations (Nyx default model with A=1, B=0, and IllustrisTNG and Illustris) are shown in Fig.~\ref{fig:tri_rho_T}. 
For each simulation, the gas is divided into four phases 
depending on the temperature and density, 
namely the \ac{WHIM}, Diffuse Ly$\alpha$, Hot Halo gas, and Condensed, 
where the cutoffs are set to be $T=10^5$~K and $\Delta$ = 120\footnote{
Here we adopt the cutoff $T = 10^5$K, and $\Delta =120$ for different gas phases following \citet{Dave2010}, and more discussion about the different cutoff can be found in \citet{Gaikwad2017}.}.
It can be seen that there exist significant dispersion in the $T$-$\Delta$ distributions of the low-$z$ IGM, i.e., the shock-heated WHIM, for all three simulations, and the fractions of the \ac{WHIM} are directly proportional to the strength of the feedback. Specifically, 
$f_\text{WHIM,Illustris} >f_\text{WHIM,IllustrisTNG}>f_\text{WHIM,Nyx}$, while Illustris implements extreme feedback, IllustrisTNG employs mild feedback and Nyx has no feedback.

In order to fit the power law relationship in the presence of dispersion in the IGM $T$-$\Delta$ distribution, we utilize the fitting procedure presented in \citetalias{Hu2022}, which fits the power law $T$-$\Delta$ relationship by binning the Diffuse \lya{} gas ($T <10^5$K and $\Delta <120$) into 20 bins based on $\log \Delta$, and applying a least squares linear fit to the mean temperatures of the gas in each bin. Here we modify the fitting range to $-0.5 < \log \Delta < 1.5$\footnote{ Such a choice of fitting range of the power law $T$-$\Delta$ relationship leads to slightly different thermal states $[T_0,\gamma]$ for the three simulations compared with those presented in previous works \citep[\citetalias{Hu2022},][]{Khaire2023}, but the difference is minor.
}, which provides a more accurate representation of the $\Delta$ range of the \lya{} absorbers at $z \sim 0.1$, which is the principal subject of this paper.

Such a fitting procedure is applied to all simulations used in this study, including all Nyx models and IllustrisTNG and Illustris simulations.  
The best-fit power law relationship based on $[T_0,\gamma]$ and the $T$-$\Delta$ distributions are illustrated in Fig.~\ref{fig:tri_rho_T}.
The figure shows that although the three simulations yield very different overall $T$-$\Delta$ distributions, their thermal state $T_0$ and $\gamma$ are however similar.

Furthermore, as described in \citetalias{Hu2022}, we vary the UV background photoionization rate, $\Gamma_{\mathHI{}}$, of the Nyx simulations in post-processing when the simulation skewers are generated, extending the parameter grid to [$\log T_0$, $\gamma$, $\log \Gamma_{\mathHI{}}$].
The value of $\Gamma_{\mathHI{}}$ we used in this study spans from $\log (\Gamma_{\mathHI{}} /\text{s}^{-1})$ = -13.834 to -12.932 in logarithmic steps of $0.075$ dex, which gives 13 values in total (see the right panel of Fig.~\ref{fig:lowz_grid}). In total, the 3D thermal grid consists of $48\times 13=624$ Nyx models.

As mentioned earlier, $T_0$ and $\gamma$ characterize the IGM thermal state at $z \gtrsim 2$, where the IGM is dominated by the power law $T$-$\Delta$ relationship. However, their efficacy as parameters for the IGM thermal state remains uncertain at $z \lesssim 1$, where a significant fraction of the gas deviates from the power-law $T$-$\Delta$ relationship due to shock heating and feedback.
In this paper, we evaluate the effectiveness of the thermal state [$T_0$,$\gamma$] as IGM parameterization at low-$z$ using the inference framework presented in \citetalias{Hu2022},
and we make use of the photoheating parameters [$A$,$B$] as an alternative set of labels as a comparison.
These labels are particularly relevant since all Nyx models used in the training procedure of our neural network, which is the major component of our inference method, are generated by varying [$A$,$B$]. This suggests that our inference framework should be capable of recovering the values of [$A$,$B$] efficiently.
Therefore, $[A,B]$ are particularly useful in the evaluation of the $[T_0,\gamma]$.
More information about the photoheating labels [$A$,$B$] is presented in Appendix.~\ref{sec:label_AB}.

\subsection{Mock spectra, Forward-modelling and VPFIT}
\label{sec:FM_VPFIT}

We follow the procedure described in \citetalias{Hu2022} 
and generate mock spectra by calculating the \lya{} optical depth ($\tau$) array along the mock LOS.
For each simulation, including all Nyx models and IllustrisTNG and Illustris, 
 an ensemble of 20,000 skewers is created.

 \begin{figure*}
 \centering
\includegraphics[width=1.0\linewidth]{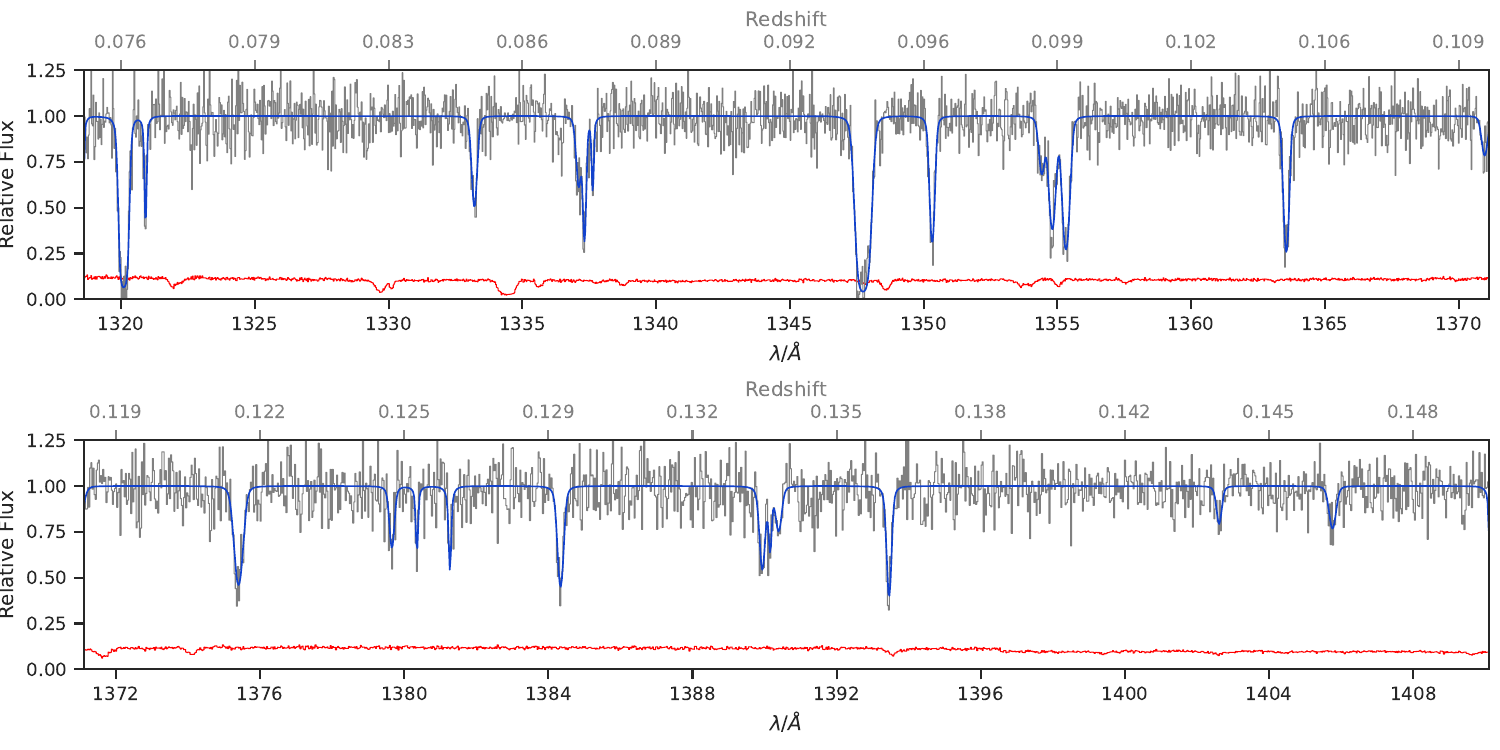}
  \caption{ One of the forward-modelled mock spectra. The simulated spectrum is shown in black, where the model spectrum determined from \vpfit{} is shown in blue, and the noise vector is plotted in red. }
  \label{fig:Nyx_model}
\end{figure*}

In this study, we include observational noise and instrumental effects to conduct our analysis under realistic conditions. We generate mock datasets with properties consistent with the \citetalias{Danforth2016} compilation of low redshift \lyaf{} spectra, which comprises 82 unique quasar spectra observed with the \ac{COS} on the \ac{HST}. 
Among them, 34 quasar spectra cover the redshift range $0.06 < z < 0.16$ with ${\rm S\slash N} > 5$ per pixel.
These spectra are used in our forward-modelling procedures, which gives a total pathlength $\Delta z_\text{ob}=2.136$.
We follow the forward-modelling procedure described in \citetalias{Hu2022}, 
in which we stitch skewers together to match the wavelength coverage of the \citetalias{Danforth2016} spectra and convolve these long skewers with the \ac{HST} \ac{COS} \ac{LSF} before finally adding Gaussian noise drawn from the noise vectors associated with a given \citetalias{Danforth2016} spectrum. 

For each simulation model, including all Nyx simulation models as well as both IllustrisTNG and Illustris simulations, we generated 1000 mock spectra with each spectrum forward-modelled based on a randomly selected \citetalias{Danforth2016} spectrum segment from redshift bin $ 0.06 < z < 0.16$, reproducing its wavelength grid, noise, and \ac{LSF}. The total pathlength for each simulation is approximately $\Delta z_{\text{tot}} \sim 60$, which ensures that the resulting \bndist{} is not biased by the choice of (mock) spectra. 

We then use VPFIT \citep[][]{vpfit}\footnote{VPFIT: \url{http://www.ast.cam.ac.uk/~rfc/vpfit.html}} to fit the \lya{} lines in our simulated spectra to obtain a set of \bn{} pairs for all of the mock datasets, following the prescription given in Hu2022.
In this paper, as is the convention in low-$z$ \lyaf{} analysis, we apply a filter for both  $b$ and $ N_{\mathHI}$,
and uses only $b$-$ N_{ \mathHI}$ pairs with $12.5 \leq \log ( N_{\mathHI} / \text{cm}^{-2}) \leq 14.5$ and $0.5 \leq \log ( b / \text{km s}^{-1}) \leq 2.5$ in our analysis \citep{schaye2000,rudie2012,Hiss2018}. 
A segment of one of the forward-modelled mock spectra is shown in Fig.~\ref{fig:Nyx_model}. The simulated spectrum is shown in grey, where the model spectrum determined from \vpfit{} is shown in blue, and the noise vector is plotted in red.

\begin{figure*}  
 \centering
    \includegraphics[width=1.0\linewidth]{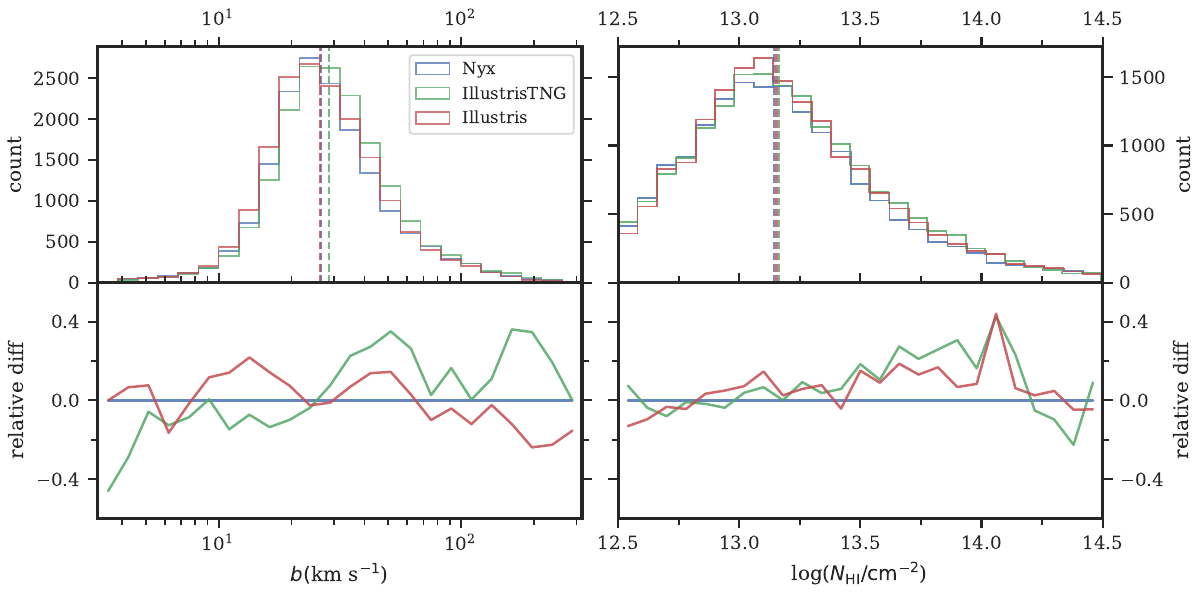}
  \caption{top: Marginalized 1D $b$ (left) and $N_{\mathHI{}}$ (right) distributions for all three simulations.  For each simulation, the \bn{} dataset is obtained by VP-fitting an ensemble of 1000 forward-modelled mock spectra. The median values for $\log b$ and $\log N_{\mathHI{}}$ are indicated by dashed vertical lines.
  bottom: The relative difference compared with Nyx simulation. The three simulations used here are \dndz{} matched.}
  \label{fig:tri_bN}
\end{figure*}

For each simulation, we apply \vpfit{} to all 1000 mock spectra to acquire a set of \bn{} parameters. 
The top panels of Fig.~\ref{fig:tri_bN} display 1D histograms of both $b$ (left) and $N_{\mathHI{}}$ (right) for all three simulations, and the bottom panels illustrate their relative differences when compared to the Nyx simulation (as discussed in \S~\ref{sec:GUV_dNdz}, the three simulations used here are \dndz{} matched). 
The median value for $\log b$ and $\log N_{\mathHI{}}$ are indicated by dashed vertical lines for each simulation. 
Notably, while the median values of both $b$ and $N_{\mathHI{}}$ are comparable across the three simulations, there are distinct differences in the distributions of both parameters across the three simulations. We also notice that the differences in the $b$ parameters are more significant across the three simulations compared with $N_{\mathHI}$. 
The three simulations are \dndz{} matched, and the relevant discussion is presented in the subsequent section.

\subsection{ Photoioniztion rate $\Gamma_{\mathHI{}}$ and \dndz}
\label{sec:GUV_dNdz}

\begin{figure}  
 \centering
    \includegraphics[width=1.0\columnwidth]{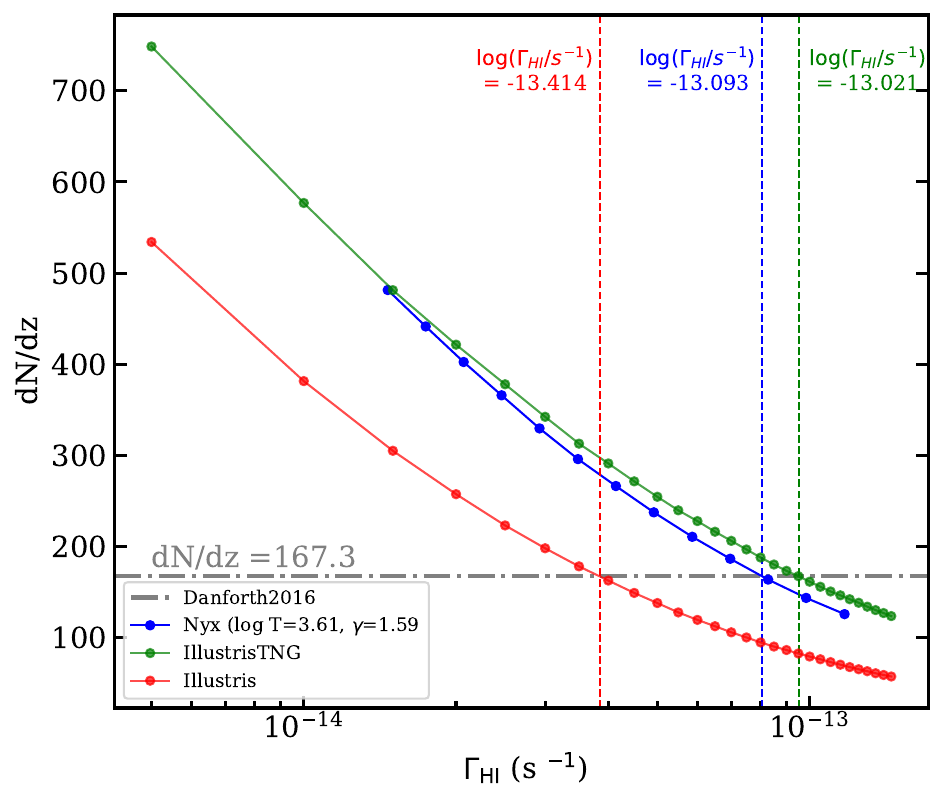}
  \caption{\dndz{} vs $\Gamma_{\mathHI{}}$ for all three simulations at $z$=0.1. Nyx (default model) is shown in blue, IllustrisTNG in green, and Illustris in red, while the observed \dndz{} calculated from \citetalias{Danforth2016} dataset for the corresponding redshift are shown in the horizontal grey dash-dotted line. The $\Gamma_{\mathHI}$ values used for each simulation to match the observed \dndz{} are indicated by vertical dashed lines with the corresponding colour.
  }
  \label{fig:dNdz}
\end{figure}

It is noteworthy that the three simulations used in this study by default have  
different UV background photoionization rates $\Gamma_{\mathHI{}}$ (for Nyx, here we are referring to the default model with $\log (T_0/\text{K})$ = 3.612 and $\gamma=1.588$.)
This is because the photoionization rate $\Gamma_{\mathHI{}}$ are tuned in post-processing across all three simulations to ensure they exhibit the same absorber density \dndz{} as the one we measured from \citetalias{Danforth2016} dataset at $z=0.1$. Specifically, we apply the aforementioned VP-fitting procedure to \citetalias{Danforth2016} spectra (segments) with $<0.06 z < 0.16$, and obtain \dndz=167.3 for absorbers within the limits $12.5 \leq \log ( N_{\mathHI} / \text{cm}^{-2}) \leq 14.5$ and $0.5 \leq \log ( b / \text{km s}^{-1}) \leq 2.5$.
Such matching of \dndz{} is analogous to the matching of the mean flux of simulations at high-$z$.
To match this \dndz{}, we tune the photoionization rate, following the prescription described in \S~\ref{sec:Grid_label}, and set {$\log (\Gamma_{\mathHI{}},  /\text{s}^{-1})$} =  -13.093, -13.021, -13.414 for Nyx, IllustrisTNG and Illustris respectively (see Fig.~\ref{fig:dNdz}).
Such mismatch in $\Gamma_{\mathHI{}}$ is caused by the degeneracy between the photoionization rate and different feedback recipes used in the simulations. Since both the UV background and feedback suppress the formation of \lya{} absorbers \citep{Khaire2023}. More specifically, the feedback heat up the IGM, converting a significant amount of the diffuse \lya{} gas into \ac{WHIM}, which reduces the \lya{} transmission caused by the neutral hydrogen \mathHI{} in the cool diffuse \lya{} gas. To this end, simulations with stronger feedback exhibit lower \dndz{} under the same $\Gamma_{\mathHI{}}$.

We measure the \dndz{} for the three simulations, including all Nyx simulation models and IllustrisTNG and Illustris, each based on its respective set of 1000 forward-modelled mock spectra.
The relationships between UV background photoionization rate and \dndz{} for all three models are shown in Fig.~\ref{fig:dNdz}, where the \dndz{} for Nyx is plotted in blue, IllustrisTNG in green, and Illustris in red,
while the \dndz{} for the D16 data at $z=0.1$ is shown as the horizontal dash-dotted grey line. 
Fig.~\ref{fig:dNdz} demonstrates that while Illustis has the strongest feedback, which causes more gas to be collisionally ionized, reducing the \lya{} absorption,
it requires the lowest $\Gamma_{\mathHI{}}$ to match the \dndz{} to the observed value, 
and IllustrisTNG, with mild feedback, has higher \dndz{} for the same UV background. 
In addition, given the diffuse \lya{} reactions in both Nyx and IllustrisTNG simulations are similar, we expected $\Gamma_{\mathHI}$ required to reproduce the observed \dndz{} to be alike, which we find to be slightly different. We are unsure of the exact reasons behind this small discrepancy, however, it might arise from the imperfection of the \vpfit{} or because of the inherent difference in the codes used for TNG and Nyx simulations, as well as their distinct implementations for various astrophysical processes.

If not otherwise specified, the three simulations used in this study, including Illustris, IllustrisTNG, and Nyx default model, are tuned to have the same \lya{} line densities, with \dndz{} $=$ 167.3, which is the same value we measured from the \citetalias{Danforth2016} dataset.

 \section{Inference Method}
  \label{sec:inference}

 \subsection{Emulating the \bn{} Distribution}
 \label{sec:emu}
 
In this work, we make use of the inference framework following \citetalias{Hu2022},
which measures the thermal state and the photoionization rate $\Gamma_{\mathHI{}}$ of the low redshift \ac{IGM} using its \bndist{} and absorber line density \dndz{}.
The \bndist{} emulator is built on \ac{DELFI}, which turns inference into a density estimation task by learning the distribution of a dataset as a function of the labels or parameters \citep{papamakarios2016, Alsing2018, papamakarios2018, Lueckmann2018, Alsing2019}.
Following \citetalias{Hu2022}, we make use of \texttt{pydelfi}, the publicly available \texttt{python} implementation of \ac{DELFI},\footnote{See https://github.com/justinalsing/pydelfi} which makes use of \ac{NDE} to learn the sampling conditional probability distribution $P(\mathbf{d} \given \boldsymbol{\theta})$ of the data summaries $\mathbf{d}$, as a function of labels/parameters $\boldsymbol\theta$, from a training set of simulated data.
Here the data summaries $\mathbf{d}$ are [$\log \text{\NHI{}}$, $\log b$], and our two sets of label parameters $\boldsymbol{\theta}$ are the thermal state [$\log T_0$, $\gamma$, $\log \Gamma_{\mathHI{}}$] and photoheating labels [$A$,$B$, $\log \Gamma_{\mathHI{}}$]. The $\Gamma_{\mathHI{}}$ grids are identical for the two sets of labels.

We generate two training datasets by labelling the \bn{} pairs obtained from our simulated spectra with the two sets of labels respectively.
We train the neural network on the summary-parameter pairs for each training dataset separately.
Our \bndist{} emulator learns the conditional probability distribution $P ( b \mathbin{,} N_{\mathHI{}} \, \given \, T_0, \gamma, \log \Gamma_{\mathHI{}})$  and  $P ( b \mathbin{,} N_{\mathHI{}} \, \given \, A, B, \log \Gamma_{\mathHI{}})$ from the corresponding training dataset. These conditional \bndist{}s are then used in our inference algorithm, where we try to find the best-fit model given the observational/mock dataset, which is described in the following section.

\subsubsection{Likelihood function}
  \label{3sec:log-likelihood}

In Bayesian inference, a likelihood $\mathcal{L}= P(\mathrm{data}|\mathrm{model})$ is used to describe the probability of observing the data for any given model.
We adopt the likelihood formalism introduced in \citetalias{Hu2022},
which is summarized as follows,
\begin{equation}
\ln \mathcal{L} = \sum_{i=1}^{n} \ln (\mu_i) - \left(\frac{\text{d} N}{\text{d} z}\right)_{\rm model}\Delta z _{\rm data}, 
\label{eq:likelihood}
\end{equation}
where $\mu_i$ is the Poisson rate of an absorber occupying a cell in the $b$-$N_{\rm HI}$ plane with area $\Delta {\rm N_{\mathHI{}}}_i\times \Delta b_i$, i.e.
\begin{equation}
\mu_{i}=\left(\frac{\text{d} N}{\text{d} z}\right)_{\rm model}\,P(b_i, N_{\mathHI{},i} \given \boldsymbol{\theta})\,\Delta { N_{\mathHI{}}}\, \Delta b\, \Delta z _{\rm data}. 
\label{eq:mu}
\end{equation}
The $P(b_i, { N_{\mathHI{}}}_i \given \boldsymbol{\theta})$ in the equation
is the probability distribution function at the point $(b_i, N_{\mathHI{},i})$
for any given model parameters $\boldsymbol{\theta}$
evaluated by the DELFI \bndist{} emulator described in \S~\ref{sec:emu}.
The $\Delta z_{\rm data}$ is the total redshift pathlength covered by the quasar
spectra from which we obtain our \bn{} dataset, and $\left({\text{d} N}\slash{\text{d} z}\right)_{\rm model}$ is the absorber density which is evaluated for any given set of parameters using a Gaussian process emulator (based on \texttt{George}, see \citealt{Ambikasaran2016}), which is also trained on our training datasets obtained from the Nyx simulation suite.

To perform our analysis under realistic conditions, all tests performed in this paper are based on mock datasets consisting of 34 forward-modelled spectra, each corresponding to one of the 34 \citetalias{Danforth2016} quasar spectra, which gives these datasets the same pathlength as the observation dataset with $\Delta z_{\text{ob}} = 2.136$.

    \begin{figure*}
 \centering
    \includegraphics[width=0.75\textwidth]{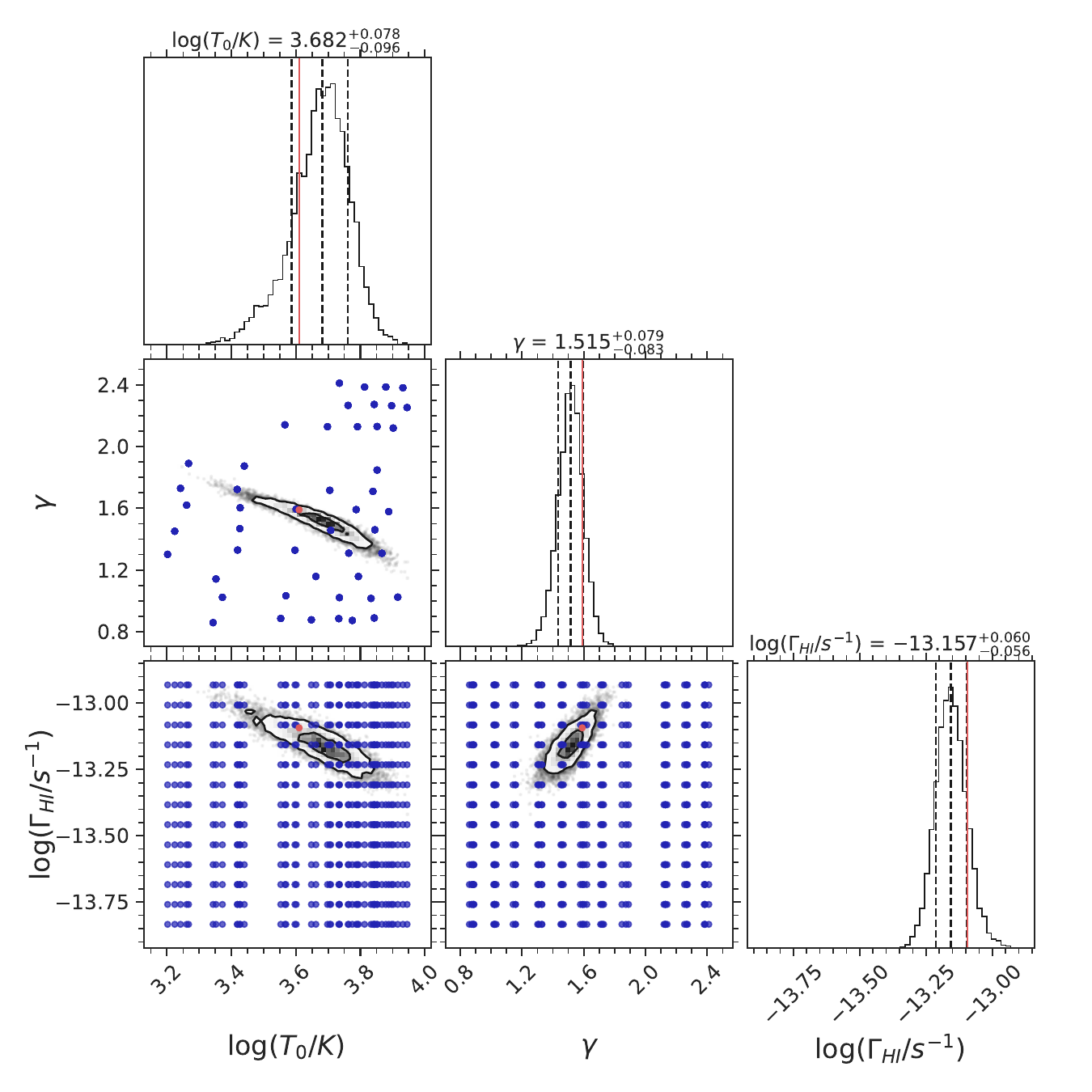}
  \caption{An example of posterior obtained by our inference method based on inference labels [$T_0$, $\gamma$, $\log \Gamma_{\mathHI}$]. Projections of the thermal grid used for generating models are shown as blue dots, while the true model is shown as red dots. The inner (outer) black contour represents the projected 2D 1(2)-sigma interval. Red lines in the marginal distributions indicate the parameters of true models, while the dashed black lines indicate the 16, 50, and 84 percentile values of the marginalized 1D posterior. The true parameters are: {$\log (T_0/\text{K})$} = 3.612 and $\gamma=1.588$, while {$\log (\Gamma_{\mathHI{}} /\text{s}^{-1})$} = -13.093. }
  \label{fig:Nyx_corner}
\end{figure*}

 \begin{figure}
 \centering
    \includegraphics[width=\columnwidth]{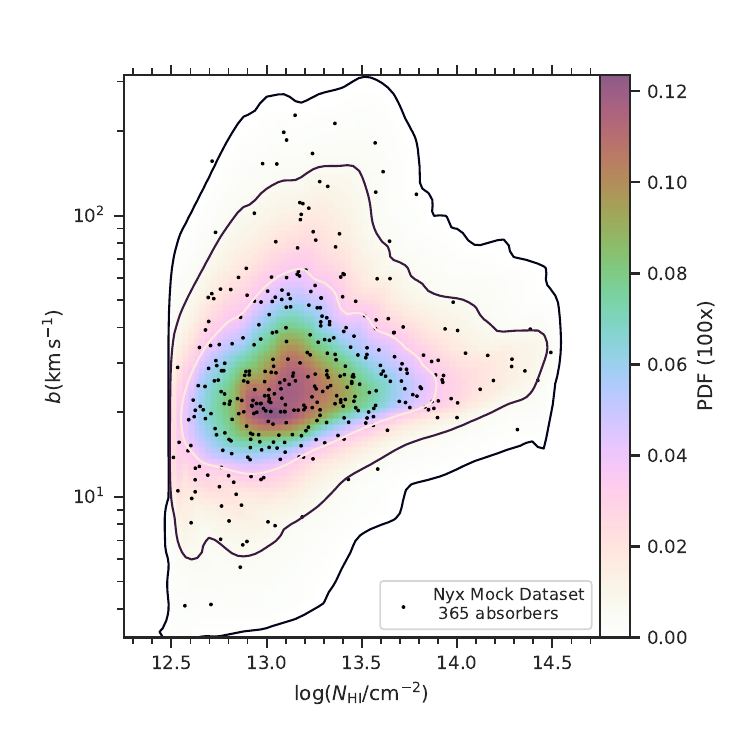}
  \caption{The colour map is the full \bndist{} recovered from the Nyx mock dataset, which is emulated by our DELFI emulator based on the best-fit parameters (median values of the marginalized MCMC posterior), where $\log (T_0/\text{K})$ = 3.682 (3.612) and $\gamma=1.515$ (1.588) and {$\log (\Gamma_{\mathHI{}} /\text{s}^{-1})$} =-13.157 (-13.093), the true parameters are given in parentheses. Black dots are the mock datasets we used in the inference. 
  The contours correspond to cumulative probabilities of 68\%, 95\% and 99.7\%.  
  For illustration purposes, the values of pdf are multiplied by 100 in the colour bar.}
  \label{fig:fit_Nyx}
\end{figure} 

An example of the MCMC posterior obtained based on the aforementioned likelihood function is given in Fig.~\ref{fig:Nyx_corner}. 
The inference is conducted using the labels [$T_0$,$\gamma$, $\log \Gamma_{\mathHI{}}$].
The posterior appears compact, with the medians of the marginalized posteriors landing close to the true parameters for all three parameters, i.e., within 1-$\sigma$ errors for marginalized 1D distributions.
The \bndist{} recovered from the mock dataset is presented in Fig.~\ref{fig:fit_Nyx},
which is emulated by our \bndist{} emulator, trained on [$T_0$, $\gamma$, $\log \Gamma_{\mathHI{}}$], based on the inferred parameters, 
i.e., median values of the marginalized 1D MCMC posterior. The plot exhibits a good match between the mock dataset (black dots) and the recovered \bndist{} (colour map).  

As a comparison to the IGM parameterization based on the thermal state, $[T_0,\gamma, \log \Gamma_{\mathHI{}}]$, the inference result derived from the same mock dataset using the photoheating labels [$A$,$B$, $\log \Gamma_{\mathHI{}}$] is given in Appendix.~\ref{sec:label_AB}.

\subsection{Inference test}
  \label{sec:inf_test}

\begin{figure}
 \centering
\includegraphics[width=0.49\textwidth]{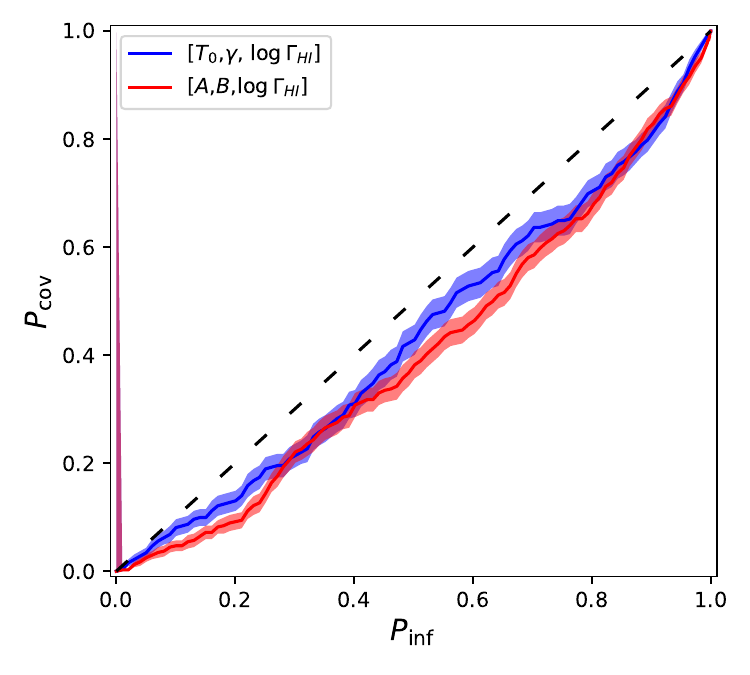}
  \caption{Coverage probability $P_\text{cov}$ for inference tests based on different labels. The x-axis stands for the inferred probability $P_\text{inf}$, and the y-axis shows the coverage probability $P_\text{cov}$ for the true parameters to fall in the contour corresponding to $P_\text{inf}$. Blue: Inference test based on the thermal state [$T_0$,$\gamma$, $\log \Gamma_{\mathHI}$], Red: Inference test based on the photoheating label [$A$,$B$,$\log \Gamma_{\mathHI}$]. 
  The shaded regions indicate the $1$-$\sigma$ error for $P_\text{cov}$.} 
  \label{fig: inference_test}
\end{figure}

An inference test is an effective method to evaluate the robustness of a given inference algorithm, which usually consists of approximations and emulation/interpolation procedures that might induce additional uncertainties, altering the error budget.
In practice, an inference test can be conducted by performing a set of realizations of the inference method using mock datasets and evaluating the robustness of the resulting posterior probability distributions, which can be 
quantified by the coverage probability $P_\text{cov}$ \citep{Prangle2014,Ziegel2014,Morrison2018,Sellentin_2019}, 
the proportion of the time that the true parameters used to generate a mock dataset are contained within the posterior contour corresponding to a certain probability level $P_\text{inf}$.
Such calculations can be performed for many different probability levels, resulting in a series of coverage probabilities. For perfect inference, this coverage probability $P_\text{cov}$ is always equal to the probability level of the chosen posterior contour $P_\text{inf}$ (shown as the black dashed line in Fig.~\ref{fig: inference_test}).

In this study, we make use of the  
inference test described in \cite{wolfson2022},
which calculates the coverage probability based on the MCMC posteriors.
Compared with the one used in \citetalias{Hu2022},  this inference test algorithm is more precise and automatically returns full coverage probabilities from 0 to 1 rather than coverage probabilities at only a few specific probability levels.

To evaluate the effectiveness of [$T_0, \gamma$, $\log \Gamma_{\mathHI{}}$] as IGM parameters for inference at low-$z$,
where the IGM $T$-$\Delta$ distribution is no longer characterized by the power-law relationship,
we perform inference tests based on different sets of labels.
We compare the result of the inference test based on labels [$T_0, \gamma$, $\log \Gamma_{\mathHI{}}$] with the one based on the photoheating labels [$A$, $B$, $\log \Gamma_{\mathHI{}}$]. As discussed in \S~\ref{sec:Grid_label}, the comparison between these two sets of labels sheds light on the efficacy of $[T_0,\gamma]$ as IGM parameters at low-$z$, where the pervasive shock heating causes significant dispersion in the IGM $T$-$\Delta$ distribution.

For each set of labels, we ran 300 realizations of our inference method, each based on a model randomly chosen from the grid. 
We exclude models that are close to the boundaries to mitigate boundary effects caused by the hard cutoff of the inference prior, which leads to the truncation of the posteriors. 
For [$T_0, \gamma$, $\log \Gamma_{\mathHI{}}$] gird, we specify 
$ 3.3 < \log (T_0/\text{K}) < 3.9$, $ 1.0 < \gamma <2.3$, $ -13.75 < \log (\Gamma_{\mathHI{}} /\text{s}^{-1})<-13.0$.
We then calculate the full coverage probabilities based on all 300 MCMC posteriors.

The results of the inference tests are shown in Fig.~\ref{fig: inference_test},
where the x-axis stands for inferred probability $P_\text{inf}$, and the y-axis shows the coverage probability, $P_\text{cov}$. The shaded regions indicate the 1-$\sigma$ error for $P_\text{cov}$, which is calculated based on the binomial distribution. The $y=x$ black-dash line represents a perfect inference test.
It can be seen that for Nyx simulations, our inference method is mildly over-confident, and the thermal state [$T_0$, $\gamma$] (blue) performs slightly better than the photoheating labels [$A$,$B$] (red), i.e., $P_\text{cov}$/$P_\text{inf}$ is closer to unity. 
Quantitatively, for inference based on the thermal state [$T_0$, $\gamma$], the 68\% contour contains the true parameters $61.2 \pm 2.8$\% of the time, and the 95\% contour contains the true parameters $90.4 \pm 1.6$\% of the time.
The results show that the [$T_0$, $\gamma$] are still robust inference labels for the IGM at low-$z$, although the shock heating alters the $T$-$\Delta$ distribution. This further suggests that the observable, i.e., the \lyaf{}, is not significantly affected by the shock heating at low-$z$.
Lastly, while the general efficacy of the inference framework remains robust, we attribute its imperfections to two primary sources: deficiencies within the neural network used in our inference algorithm, and the boundary effects caused by the truncation of the posteriors when hitting the boundary.

\subsection{Inference results for IllustrisTNG and Illustris}
  \label{sec:inf_ILL}

In this section, we employ the IllustrisTNG and Illustris simulations as mock observational data to explore the impacts of feedback, mainly AGN feedback, on the IGM thermal state $[T_0,\gamma]$. 
More specifically, we evaluate the robustness of our inference method, built on the Nyx thermal grid without galaxy formation and feedback, when applied to observational data derived from a (mock) Universe with substantial feedback associated with galaxy formation and AGN activities. 
The investigation is broken down into two separate inquiries. First, it explores the extent to which feedback associated with galaxy formation and AGN activities impacts the \lyaf{}. Second, it investigates how, given the presence of these effects, the feedback influences the inferred parameters $[T_0,\gamma]$.

\begin{figure*}
 \centering
    \includegraphics[width=0.49\textwidth]{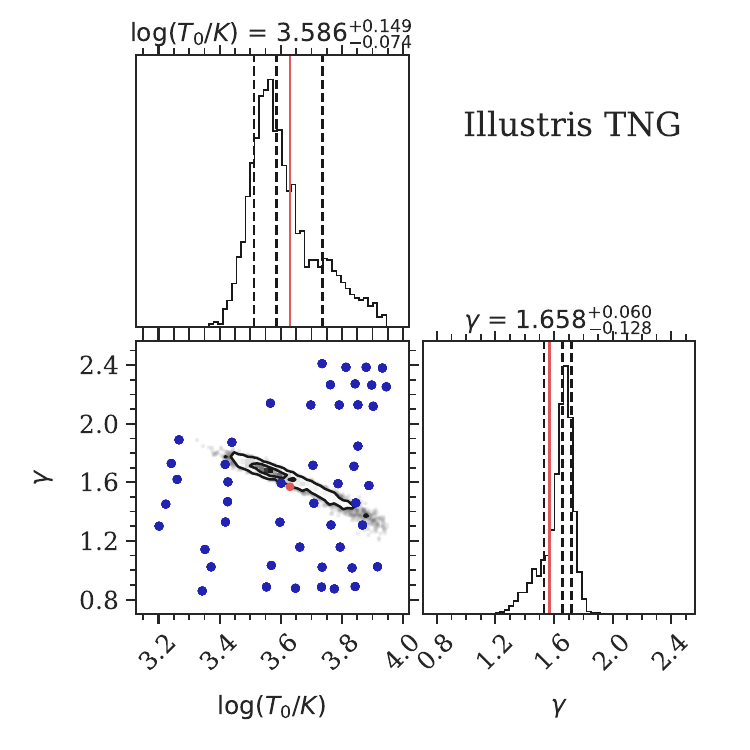}
    \includegraphics[width=0.49\textwidth]{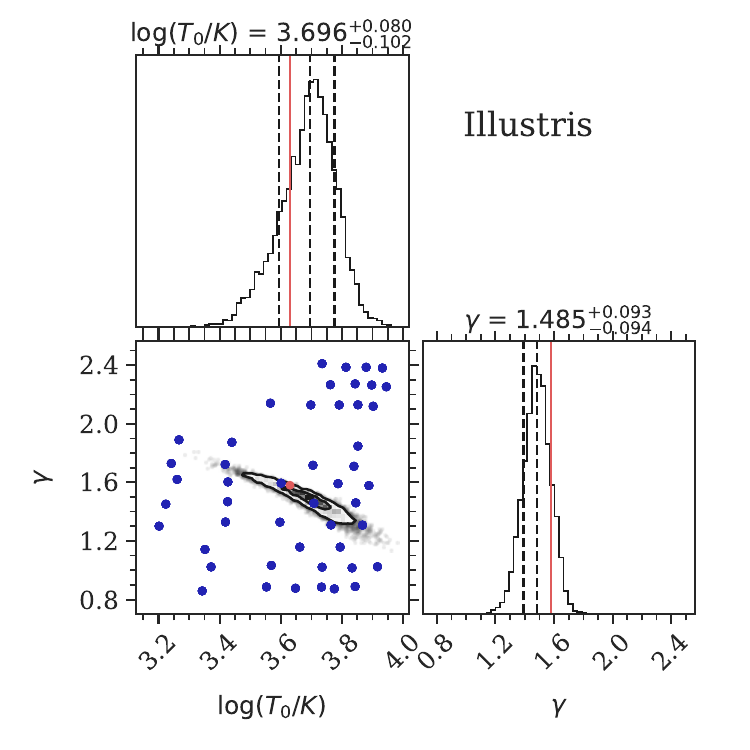}
  \caption{ Posteriors obtained by using IllustrisTNG (left) and Illustris (right) as mock observational data in our inference method. Projections of the thermal grid used for generating models are shown as blue dots. The inner (outer) black contour represents the projected 2D 1(2)-sigma interval. The true parameters for the simulations, obtained by fitting the $T$-$\Delta$ distributions of the simulations, are indicated by the red dot (lines) in the (marginal) distributions, while the dashed black lines indicate the 16, 50, and 84 percentile values of the marginalized 1D posterior. }
  \label{fig:corner_ILL}
\end{figure*}

\begin{figure*}
 \centering
     \includegraphics[width=0.49\textwidth]{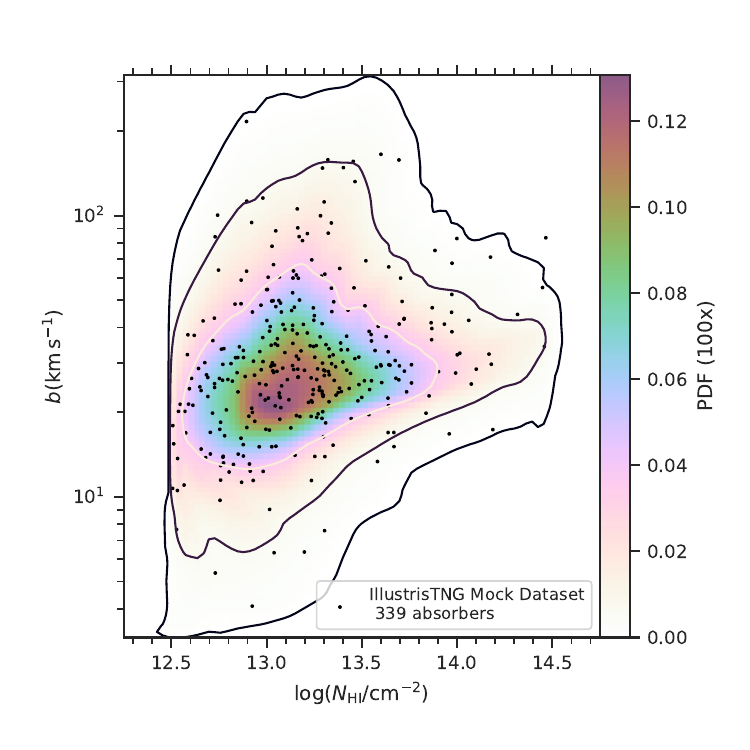}
    \includegraphics[width=0.49\textwidth]{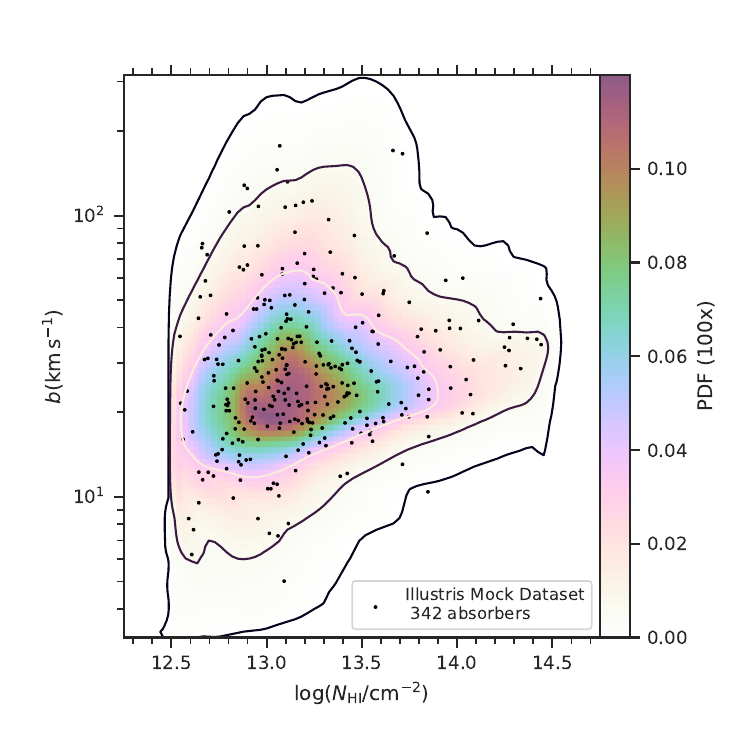}
  \caption{  Joint \bndist{}s recovered from the inference results for IllustrisTNG (left) and Illustris (right) simulations, emulated by our DELFI emulator based on the median values of the marginalized MCMC posterior. Black dots are the mock datasets we used in the inference. The contours correspond to cumulative probabilities of 68\%, 95\% and 99.7\%.  For illustration purposes, the values of the pdf are multiplied by 100 in the colour bar. }
  \label{fig:fit_ILL}
\end{figure*}

Following the forward-modelling prescription described in \S~\ref{sec:FM_VPFIT},
we generate mock datasets with $\Delta z$=2.136, the pathlength of \citetalias{Danforth2016} dataset at $z=0.1$,  
for both simulations (see Fig.~\ref{fig:fit_ILL}),
and run our inference method on each dataset. 
As discussed in \S~\ref{sec:GUV_dNdz}, a degeneracy exists between the strength of the AGN feedback implemented in the simulations and the UV background photoionization rate $\Gamma_{\mathHI{}}$, both of which suppress the abundance of absorbers, hence reducing the \dndz~\citep[see][for more details]{Khaire2023}. 
Given that our inference method primarily derives the photoionization rate $\Gamma_{\mathHI{}}$ based on the \dndz, the resulting $\Gamma_{\mathHI{}}$ always aligns with the value that generates the equivalent \dndz{} in the Nyx simulation (see \S~\ref{sec:GUV_dNdz}).
Since here we use IllustrisTNG and Illustris simulations with their \dndz{} matched to the \citetalias{Danforth2016} low-$z$ dataset, the inferred $\Gamma_{\mathHI{}}$ always disagrees with the true values used to generate the IllustrisTNG and Illustris simulations.
To this end, we conduct our inference test in 2D without considering the accuracy with which we recover the photoionization rate $\Gamma_{\mathHI{}}$. Posterior distributions for the thermal parameters obtained from our inference applied to Illustris and IllustrisTNG are shown in  Fig.~\ref{fig:corner_ILL}, where we have marginalized over $\Gamma_{\mathHI{}}$. 
For these two mock datasets, we infer that  [$\log (T_0/\text{K})$, $\gamma$] = $[3.586^{+0.149}_{-0.074},1.658^{+0.060}_{-0.128}]$  
for IllustrisTNG ([3.627,1.593]), and [$\log (T_0/\text{K})$, $\gamma$] = $[3.696^{+0.080}_{-0.102}, 1.485^{+0.093}_{-0.094}]$ for Illustris ([3.633,1.577]), while the true parameters for the two simulations, $[T_0,\gamma]_\text{fit}$, are given in parentheses respectively.

It can be seen that the true parameters $[T_0,\gamma]_\text{fit}$, obtained by fitting the $T$-$\Delta$ distributions of the simulations, are within 1-$\sigma$ errors (1D marginalized) for both simulations, and the 1-$\sigma$ errors for both the IllustrisTNG and Illustris simulations are slightly larger than those for Nyx simulations, which is caused by the intrinsic difference between Nyx, IllustrisTNG and Illustris simulations, 
where the latter two are based on completely different hydrodynamic codes.
In Fig.~\ref{fig:fit_ILL}, we present both the mock datasets used for inference and the \bndist{}s emulated based on the inference results. The plots highlight strong agreement between the emulated \bndist{}s and the respective mock dataset for each simulation.

\begin{figure*}
 \centering
     \includegraphics[width=0.45\textwidth]{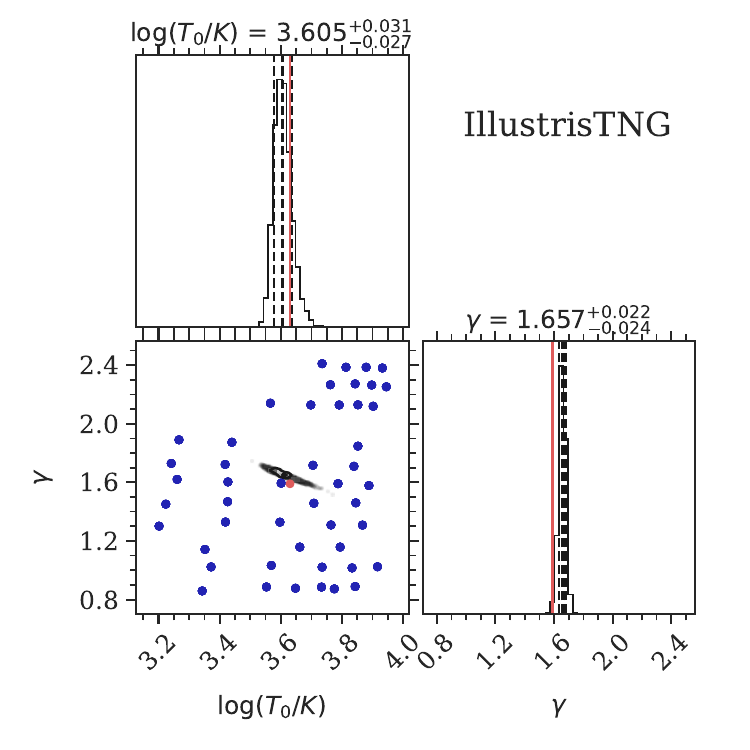}
    \includegraphics[width=0.45\textwidth]{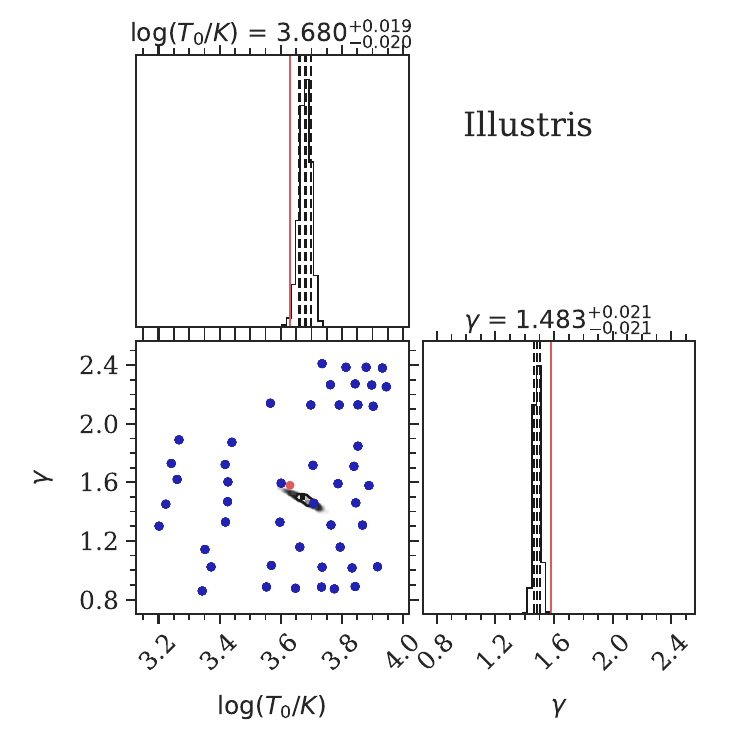}
  \caption{ Corner plots for IllustrisTNG (left) and Illustris (right), based on the larger mock dataset, with $\Delta z=42.72$, corresponding 20 times the observational dataset. Projections of the thermal grid used for generating models are shown as blue dots. The inner (outer) black contour represents the projected 2D 1(2)-sigma interval. The true parameters for the simulations, obtained by fitting the $T$-$\Delta$ distributions of the simulations, are indicated by the red dot (lines) in the (marginal) distributions, while the dashed black lines indicate the 16, 50, and 84 percentile values of the posterior. }
  \label{fig:ILL_optimized_corner}
\end{figure*}

Nevertheless, it is worth mentioning that the inferred thermal states for IllustrisTNG and Illustris presented above are based on realistic conditions, with total pathlength $\Delta z= 2.136$. Such a small $\Delta z$  makes the inference result vulnerable to randomness induced by the selection of mock datasets. 
To address this issue, here we conduct our inference on IllustrisTNG and Illustris simulations, using datasets with much larger pathlength, specifically with $\Delta z=42.47$, which is 20 times the size of the observational dataset. The inference results yield [$\log (T_0/\text{K})$, $\gamma$] = $[3.605^{+0.031}_{-0.027},1.657^{+0.022}_{-0.024}]$ for IllustrisTNG ([3.627,1.59]), and [$\log (T_0/\text{K})$, $\gamma$] = $[3.680^{+0.019}_{-0.020}, 1.483^{+0.021}_{-0.021}]$ for Illustris ([3.633,1.58]), while the true parameters for the two simulations, $[T_0, \gamma]_\text{inf}$, are given in parentheses.
The resulting corner plots are presented in Fig.~\ref{fig:ILL_optimized_corner}. 
These results are used as our inferred thermal states $[T_0, \gamma]_\text{inf}$ for IllustrisTNG and Illustris simulations in the following part of this study.  
It is noticeable that the inferred $T_0$ for Illustris is higher than the true value with an error $\Delta \log(T_0/\text{K}) = 0.047$ dex, while the $\gamma$ is below the true value, with $\Delta \gamma = -0.094$. 
For IllustrisTNG, the offsets between the $[T_0, \gamma]_\text{inf}$ and $[T_0, \gamma]_\text{fit}$ are smaller, with  $\Delta \log(T_0/\text{K}) = -0.022$ dex,  $\Delta \gamma = 0.064$.
We notice that these offsets are smaller than the typical inference precision obtained based on realistic datasets, as shown in Fig.~\ref{fig:Nyx_corner} and Fig.~\ref{fig:corner_ILL}, which report the marginalized 1D 1-$\sigma$ error in $\log T_0$, $\sigma_{\log T_0}$, $\sim 0.1 $ dex and the marginalized 1D 1-$\sigma$ error in $\gamma$, $\sigma_{\gamma}$, $ \sim 0.1$.
For both simulations we observe the offsets $ \Delta \log T_0 \lesssim $ $0.5 \sigma_{\log T_0}$, and $\Delta \gamma \lesssim \sigma_ {\gamma}$.

\begin{figure*}
 \centering
\includegraphics[width=0.49\textwidth]{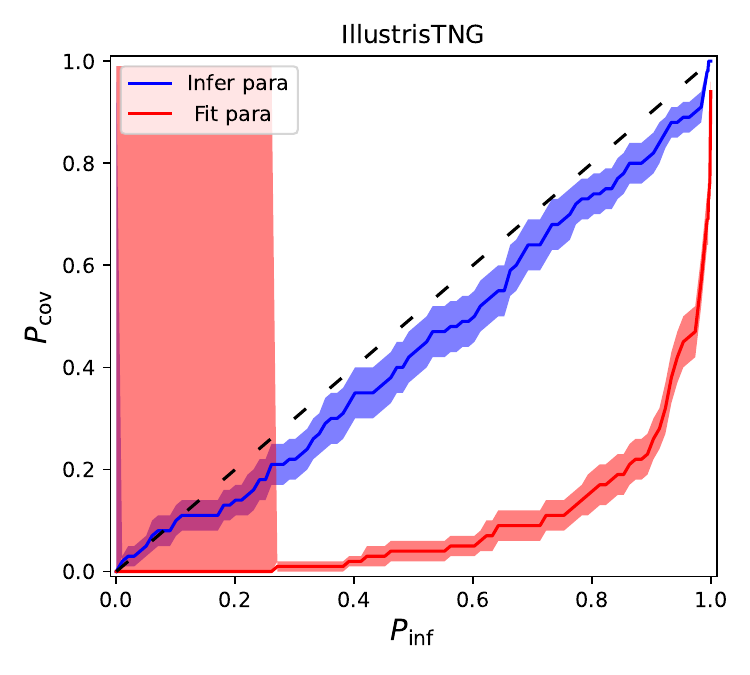}
\includegraphics[width=0.49\textwidth]{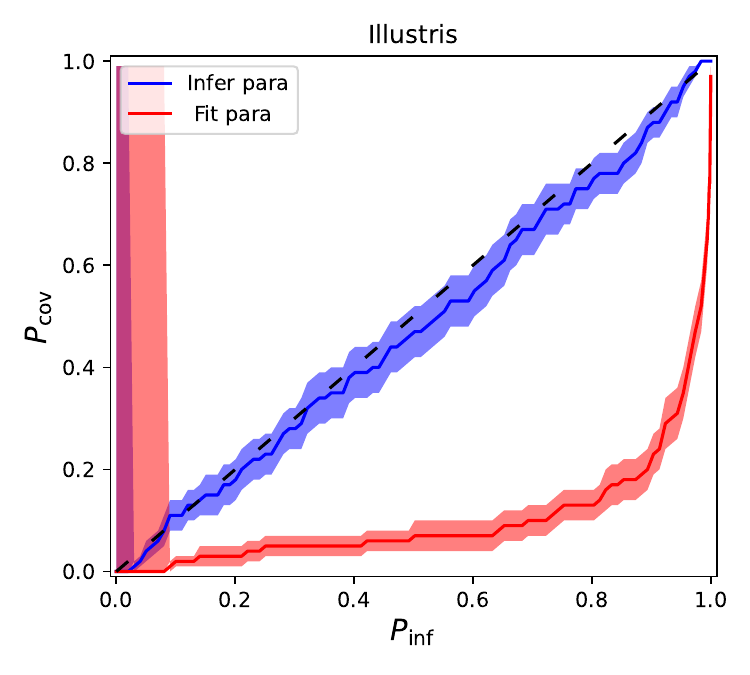}
  \caption{Marginalized coverage probability $P_\text{inf}$ for inference tests using IllustrisTNG (left) and Illustris (right) simulations as mock observational data. The x-axis stands for the inferred probability $P_\text{inf}$, and the y-axis shows the probability for the parameters of the true model to fall in the contour corresponding to $P_\text{cov}$. 
  The shaded regions indicate the $1-\sigma$ error for $P_\text{cov}$.
   Inference tests with the true parameters set by $[T_0, \gamma]_\text{inf}$ are shown in blue, while inference tests with the true parameters set by  $[T_0, \gamma]_\text{fit}$ are shown in red. } 
  \label{fig: inference_ILL}
\end{figure*}

To check the robustness of these results, we use the IllustrisTNG and Illustris simulations as mock observational data and perform inference tests using two different sets of 'true parameters': the $[T_0, \gamma]_\text{fit}$ obtained from our power law fits the $\Delta-T$ distribution of the simulations (see Fig.~\ref{fig:tri_rho_T}), 
and the $[T_0, \gamma]_\text{inf}$ given by our inference method when applied to an extremely large mock dataset, as described above. 
Given that the inferred $\Gamma_{\mathHI{}}$ for both IllustrisTNG and Illustris simulations consistently deviates from the actual values, owing to the previously mentioned degeneracy between the photoheating rate and feedback strength, any inference tests incorporating the $\Gamma_{\mathHI{}}$ from these two simulations will surely fail.
To this end, we focus on the inference results on the $T_0$-$\gamma$ plane and
conduct marginalized inference tests by marginalizing the posteriors
over the $\Gamma_{\mathHI{}}$, in which 2D marginalized contours levels are modelled by Gaussian mixture models. 
For each simulation, we run 100 realizations on each set of 'true parameters', 
and run inference tests on the obtained posteriors. 
The results are shown in  Fig.~\ref{fig: inference_ILL}, indicating that our inference method is over-confident for both sets of 'true parameters'. While the inference method is not able to recover the thermal state $[T_0, \gamma]_\text{fit}$, the thermal state $[T_0, \gamma]_\text{inf}$ significantly improves the outcome of the inference test. These results suggest that our inference method is able to robustly recover the  $[T_0, \gamma]$ with small biases, for simulations that include feedback mechanisms.

The inference tests imply that there exist offsets for the inferred parameters $[T_0, \gamma]_\text{inf}$ for IllustrisTNG and Illustris, where  $\Delta \log(T_0/\text{K}) = -0.022$ dex, $\Delta \gamma = 0.064$ for IllustrisTNG and $\Delta \log(T_0/\text{K}) = 0.047$ dex, $\Delta \gamma = -0.094$ for Illustris. However, these offset are insignificant, with $\Delta \log T_0 \lesssim $ $0.5 \sigma_{\log T_0}$, and $\Delta \gamma \lesssim \sigma_ {\gamma}$.
However, it is unclear whether the observed differences between $[T_0, \gamma]_\text{inf}$ and $[T_0, \gamma]_\text{fit}$ can be attributable to the intrinsic difference between the Nyx, IllustrisTNG, and Illustris simulations, or if they arise from potential degeneracy between the IGM thermal state $[T_0,\gamma]$ and the feedback mechanism implemented in the simulation. 
Nevertheless, the latter hypothesis seems to contrast with the results based on the various statistics of the low-$z$ \lyaf{}  presented in \citet{Khaire2023}, which suggests that the impacts from different feedback models are not distinguishable via the \lyaf{} under realistic scenarios. 
To further explore this problem, we examine the physical properties of low -$z$ \lya{} absorbers in the following section.

\section{Low-$z$ \lya{} Forests and Simulated Absorbers}
\label{sec:lowz_lyaf}

\subsection{Identifying the simulated \lya{} absorbers}
\label{sec:raw_peaks}
     \begin{figure*}
 \centering
    \includegraphics[width=0.49\linewidth]{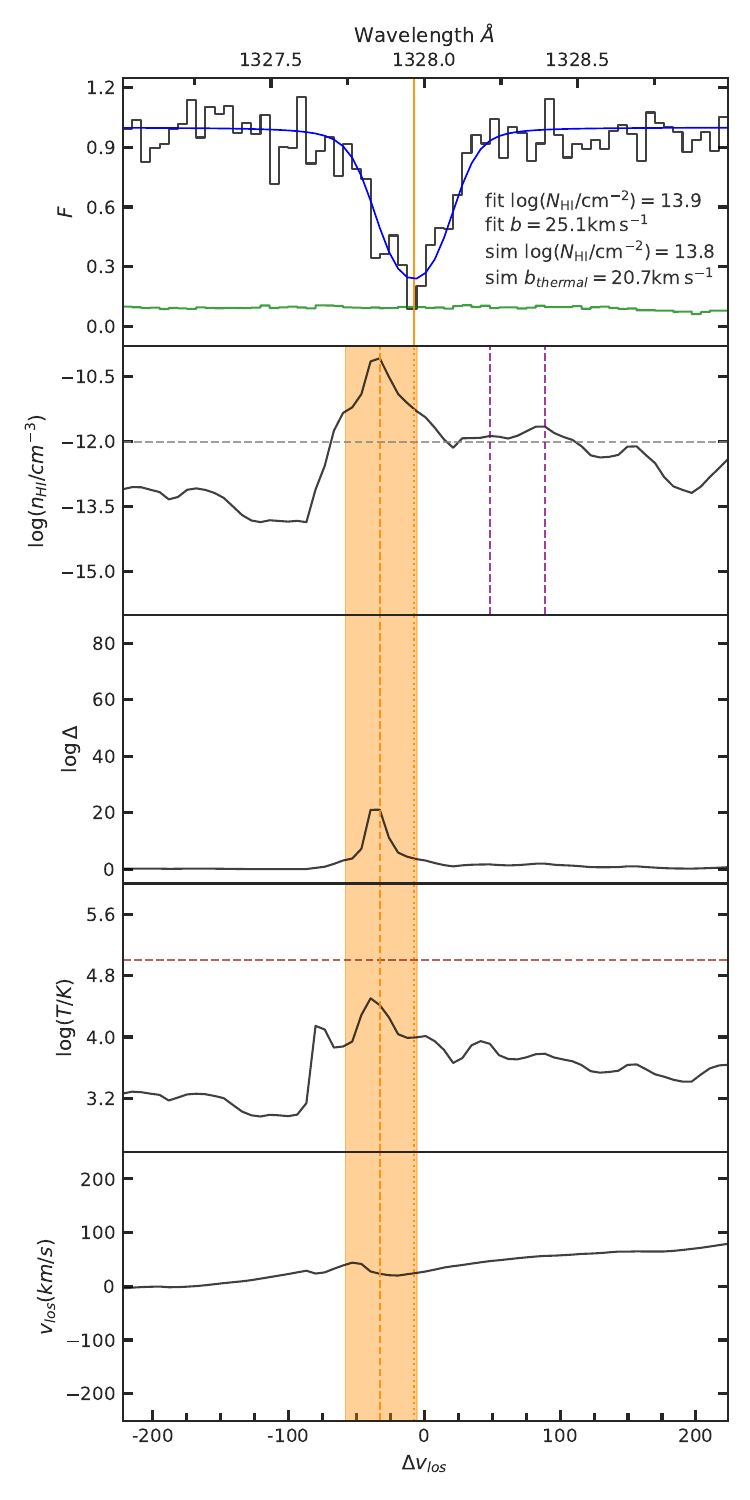}
    \includegraphics[width=0.49\linewidth]{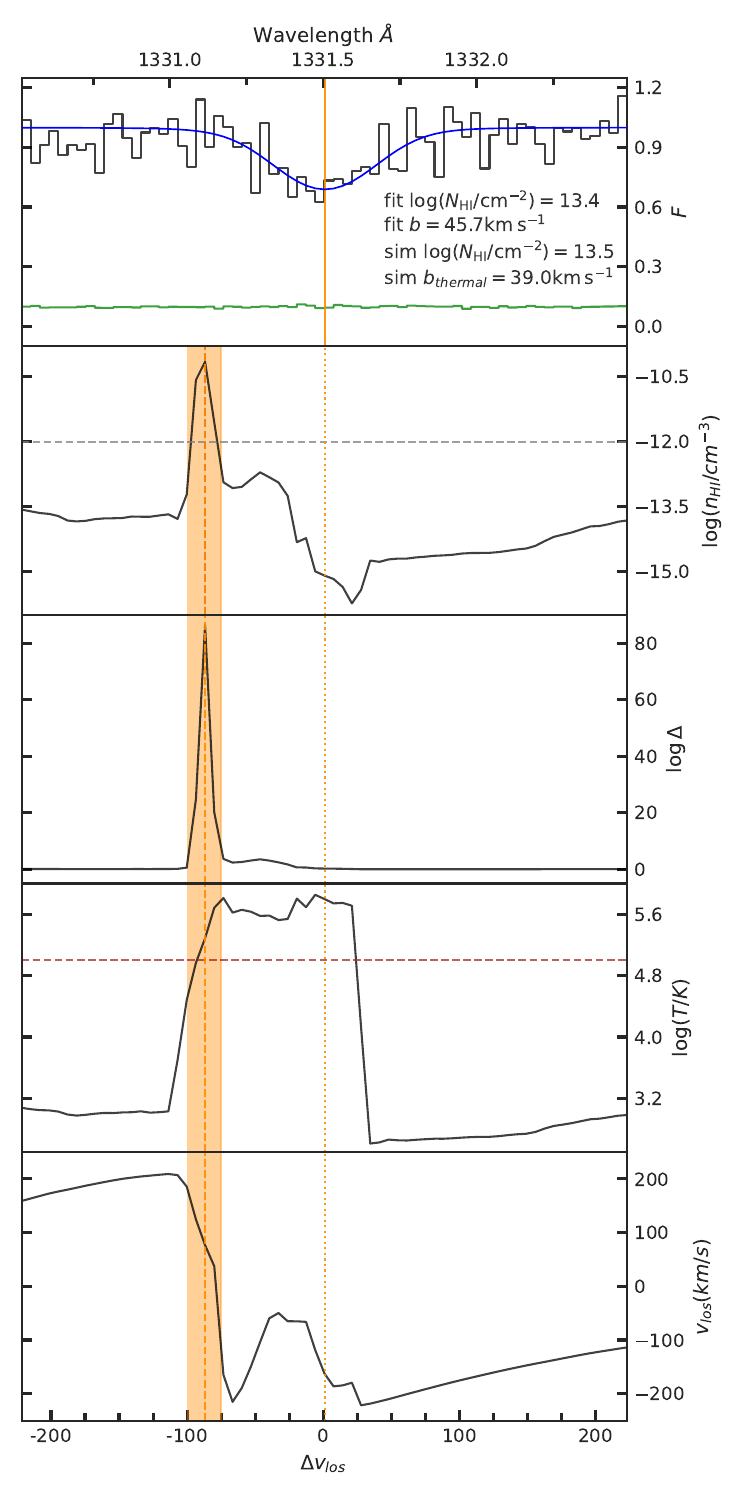}
  \caption{ 
Illustration of a segment of one of the forward-modelled mock spectra (top panel) with the absorption lines detected by VPFIT and the corresponding skewer. The top panel depicts the flux (black), noise vector (green), and the model fitted by \vpfit{} (blue). The central wavelength of \lya{} lines identified by \vpfit{} are indicated by orange vertical lines, and the corresponding simulated absorbers are indicated by orange dashed lines in the second panel (and all other panels below). 
The $\log {N_{\mathHI{}}}_\text{,fit}$, $b_\text{fit}$ reported by \vpfit{} are given in the annotation, together with the $\log  {N_{\mathHI{}}}_\text{,sim}$, $b_\text{thermal}$ calculated based on the simulation. 
The second panel depicts the neutral hydrogen density $n_{\mathHI{}}$, 
while the shaded regions represent the identified \lya{} absorbers along LOS, 
which are used as the integral ranges while computing the ${N_{\mathHI{}}}_\text{,sim}$. 
The orange vertical dashed lines show the $n_{\mathHI{}}$ peaks of the confirmed simulated \lya{} absorbers, while the purple vertical dashed lines show the potential simulated \lya{} absorbers that do not cause detectable \lya{} lines.
The grey horizontal dashed line represents the minimal H~{\sc i} peak density, $n_{\mathHI, \text{min}} =10^{-12}$ $\text{cm}^{-3}$. 
The third, fourth and fifth panels show the overdensity $\Delta$, temperature $T$ and LOS velocity $v_\text{los}$.
The brown horizontal dashed line in the fourth panel stands for $T=10^5$ K.
left: A \lya{} absorbers in the diffuse \lya{} phase.  right: A \lya{} absorbers in the WHIM phase. }
  \label{fig:Nyx_lines}
\end{figure*} 

To understand whether the low-$z$ \lyaf{} effectively probes the WHIM, 
we attempt to identify the simulated \lya{} absorbers, i.e., the $n_{\mathHI}$ peaks in the simulation skewers, that give rise to the \lya{} lines detected in the mock spectra.
This approach allows us to directly examine the physical properties ($T$, $\Delta$, and $n_{\mathHI}$) of these simulated \lya{} absorbers and draw a direct correspondence between them and the line parameters (\bn{}) of their corresponding \lya{} lines detected in the mock spectra.
In this section, we chose to focus on the simulated \lya{} absorbers in the Nyx simulation at $z=0.1$ (with default thermal history, i.e., $T_0 =3.612$, and $\gamma =1.588$ at $z=0.1$).
For clarification, within the context of this study, the terms 'simulated \lya{} absorbers' or simply 'simulated absorbers' are used to denote the  $n_{\mathHI}$ peaks that give rise to the \lya{} absorption lines in the mock spectra detected by \vpfit{}.

Our approach for identifying simulated \lya{} absorbers works as follows. 
Firstly, we include the physical properties, including temperature $T$, over-density $\Delta$, 
velocity along LOS $v_\text{los}$, and the neutral fraction $x_\mathHI{}$ in our skewers and stitch them in the forward-modelling procedure (see \S~\ref{sec:FM_VPFIT}).
We interpolate the stitched skewer onto the forward-modelled wavelength grid,
and calculated the neutral hydrogen density $n_{\mathHI{}}$ for each simulation cell, based on the neutral fraction $x_{\mathHI{}}$, over-density $\Delta$, and the mean hydrogen 
density ${\bar{n}}_\ion{H}{}$.
Subsequently, we scan the stitched skewers (in real space) for $n_{\mathHI{}}$ peaks, and classify these with $n_{\mathHI{}} > 10^{-12}$ $\text{cm}^{-3}$ as potential simulated \lya{} absorbers.
The minimal peak H~{\sc i}
density is derived from both the minimal H~{\sc i} column density for the detected lines $N_{\mathHI,\text{min}}=10^{12.5} \text{cm}^{-2}$ (see \S~\ref{sec:FM_VPFIT}) and the 
maximal length for simulated absorbers $l_\text{abs, max} =0.5$ Mpc/h, which is consistent with previous studies that attempt to characterize the structures giving rise to the \lyaf{} at $z=0.1$ 
\citep{,Bolton2021, Tillman2023}.
Given these two parameters, the requisite minimum H~{\sc i} peak density for simulated absorbers to yield observable \lya{} absorption lines is computed as $n_{\mathHI, \text{min}} = \frac{N_{\mathHI,\text{min}}}{l_\text{abs, max}} =  \frac{10^{12.5} \text{cm}^{-2}} {1 \text{Mpc/h}} \sim 10^{-12}$~$\text{cm}^{-3}$, which effectively filters
out $n_{\mathHI}$ peaks that give rise to \lya{} absorption lines below our sensitivity.
We then determine the physical size for each potential simulated absorber along the LOS, $l_{\text{abs}}$, using a threshold at which $n_{\mathHI{}}$ drops below 1\% of its peak value, while restricting the maximal size to be $l_{\text{abs, max}}=$ 0.5 Mpc/h. 
We calculate the H~{\sc i} column densities of the simulated \lya{} absorbers, $N_{\mathHI\text{, sim}}$, 
by integrating the $n_{\mathHI}$ over the ranges set by aforementioned threshold. We observed that the resulting $N_{\mathHI{}}$ is not particularly sensitive to $l_{\text{abs}}$, because the $n_{\mathHI{}}$ peak is so narrow that the majority of the neutral hydrogen comes from the peak region (see Fig.~\ref{fig:Nyx_lines}).

After identifying the potential \lya{} absorbers, we extract their LOS velocity from the simulation cells, 
and compute the central wavelength of the expected absorption lines in redshift space, accounting for the redshift caused by its LOS velocity. 
For each anticipated absorption line originating from an $n_{\mathHI{}}$ peak, 
we check whether its central wavelength lies within $\pm$ 50km/s 
of the central wavelength of any \lya{} lines detected in the mock spectrum.
If so, we confirm the identification of a simulated \lya{} absorber, and take the $T$ and $\Delta$ at the $n_{\mathHI{}}$ peak as its values, which is valid since the $n_{\mathHI{}}$ peak is so narrow that the majority of the $N_{\mathHI{}}$ comes from the region close to the peak.
While theoretically, the \lya{} lines are expected to be caused by multiple $n_{\mathHI{}}$ peaks in real space \citep{Garzilli2015}, we discover that at $z=0.1$, each \lya{} line detected in the mock spectra predominantly originates from one single $n_{\mathHI{}}$ peak in the simulation.
As such, we only consider the $n_{\mathHI{}}$ peak with the highest $n_{\mathHI{}}$ value if multiple $n_{\mathHI{}}$ peaks contribute
to the same detected absorption line.

Fig.~\ref{fig:Nyx_lines} showcases examples of the simulated \lya{} absorbers, alongside their corresponding absorption lines in the mock spectra and the related simulation skewers.
The top panel depicts the flux (black), noise vector (green), and the model fitted by \vpfit{} (blue). The central wavelength of \lya{} lines identified by \vpfit{} are indicated by orange vertical lines, and the corresponding simulated absorbers are indicated by orange dashed lines in the second panel (and all other panels below). 
The $\log {N_{\mathHI{}}}_\text{,fit}$, $b_\text{fit}$ reported by \vpfit{} are given in the annotation, together with the $\log  {N_{\mathHI{}}}_\text{,sim}$, $b_\text{thermal}$ calculated based on the simulation, whereas the $b_\text{thermal}=(2 k T / m_{\mathHI})^{1/2}$ is the thermal component of the $b$-parameters computed based on the $T$ of the simulated \lya{} absorbers (see eq.~\ref{eq:b-para} ).
The second panel depicts the neutral hydrogen density $n_{\mathHI{}}$, 
while the shaded regions represent the identified \lya{} absorbers along LOS, 
which are used as the integral ranges while computing the ${N_{\mathHI{}}}_\text{,sim}$. 
The orange vertical dashed lines show the $n_{\mathHI{}}$ peaks of the confirmed simulated \lya{} absorbers, while the purple vertical dashed lines show the potential simulated \lya{} absorbers that do not cause detectable \lya{} lines.
The grey horizontal dashed line represents the minimal H~{\sc i} peak density, $n_{\mathHI, \text{min}} =10^{-12}$~$\text{cm}^{-3}$. 
The third, fourth and fifth panels show the overdensity $\Delta$, temperature $\log T$ and LOS velocity $v_\text{los}$ (black solid lines).
The brown horizontal dashed line in the fourth panel stands for $T=10^5$ K, which divides the cool diffuse \lya{} gas and the WHIM.
The left panel shows a simulated \lya{} absorber in the diffuse \lya{} phase, while the right left panel shows a simulated \lya{} absorber arising from the WHIM phase.

We perform the identification procedure for all 1000 mock spectra, discovering 34011 potential simulated \lya{} absorbers,i.e.,  $n_{\mathHI{}}$ peaks, among which 10510 are identified as simulated \lya{} absorbers and matched to their respective absorption lines identified by VPFIT.
The discrepancy between potential and confirmed \lya{} absorbers is due to the inclusion of minor $n_{\mathHI{}}$ peaks, that are too weak to cause any detectable \lya{} line, which is indicated by purple vertical lines in the left panel of Fig.~\ref{fig:Nyx_lines}. 
Lastly, approximately 2\% of the lines detected by VPFIT could not be matched to any simulated \lya{} absorber.
These anomalies could potentially result from false identification of the \vpfit{} induced by noise. 
Nonetheless, given the rarity of these cases, omitting them should not influence our statistical results or conclusions.

\begin{figure*}
 \centering
    \includegraphics[width=1.00\linewidth]{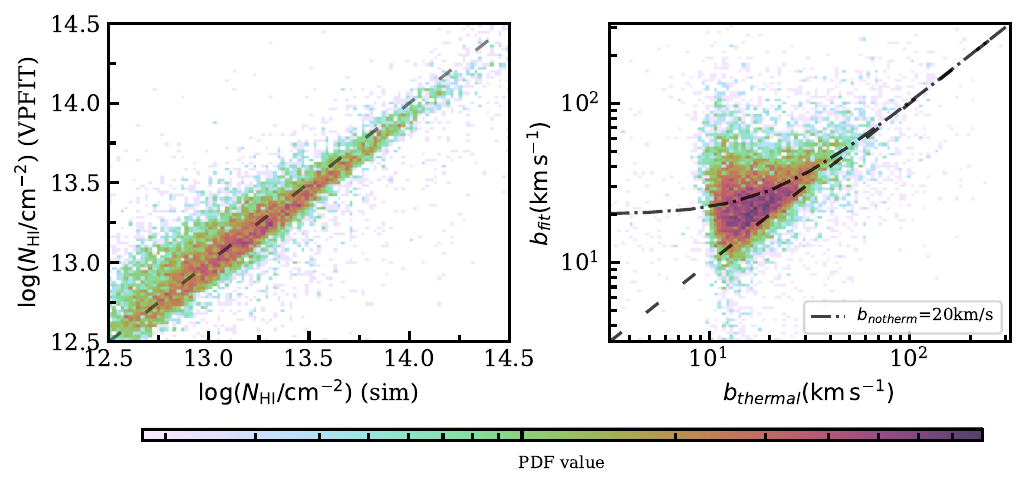}
  \caption{ The observed variables  ${N_{\mathHI{}}}_\text{,fit}$, $b_\text{fit}$ fitted by VPFIT compared with the physical quantities  ${N_{\mathHI{}}}_\text{,sim}$ and $b_\text{thermal}$ of the simulated absorbers identified in the simulation skewers, where the  ${N_{\mathHI{}}}_\text{,sim}$ is calculated by integrating the $n_{\mathHI{}}$ of the absorbers along the LOS, and the  $b_\text{thermal}$ is computed by assuming the broadening of the \lya{} lines are pure thermal.
left: ${N_{\mathHI{}}}_\text{fit}$ vs ${N_{\mathHI{}}}_\text{,sim}$.
right: $b_\text{fit}$ vs $b_\text{thermal}$.
The dash-dot line in the right panel represents the $b$-parameter resulting from the combination of the thermal component $b_\text{thermal}$ and a turbulence in the IGM with $ b_\text{notherm}=20$ km/s.
  }
  \label{fig:b_NHI_sim_fit}
\end{figure*}

To validate our identification method, we compare the observed line parameters, reported by \vpfit{}, with the values calculated from the simulation.
In Fig.~\ref{fig:b_NHI_sim_fit}, we showcase the  ${N_{\mathHI{}}}_\text{,fit}$ (left) and $b_\text{fit}$ (right) for all \lya{} lines fitted by VPFIT, compared with the ${N_{\mathHI{}}}_\text{,sim}$ and $b_\text{thermal}$ respectively, both calculated from the corresponding simulated \lya{} absorbers identified in the Nyx simulation.
The left panel indicates a strong correlation between the fitted ${N_{\mathHI{}}}_\text{,fit}$ and the ${N_{\mathHI{}}}_\text{,sim}$ calculated from the simulation, implying that the $n_{\mathHI{}}$ peaks identified by our method are indeed the simulated \lya{} absorbers responsible for the \lya{} lines detected in the mock spectra.
The right panel demonstrates that the bulk of $b_\text{fit}$ lies above the dashed line representing $b_\text{fit} = b_\text{thermal}$.
This result aligns with the nature of the $b$-parameter, as given by
\begin{equation}
    b = \sqrt{b_\text{thermal}^2 + b_\text{notherm}^2} ,
    \label{eq:b-para}
\end{equation}
where the $b_\text{notherm}$ is the non-thermal component of the $b$-parameter resulting from combinations of Hubble flow, peculiar velocities and turbulence in the IGM. 
Eq.~\ref{eq:b-para} demonstrates that the $b_\text{thermal}$ is the lower limit of the $b$-parameter, which corresponds to the lower right cutoff of the \bndist{} \citep[see the colour maps in Fig.~\ref{fig:fit_Nyx} and Fig.~\ref{fig:fit_ILL} as examples. More discussions on this topic can be found in][and \citetalias{Hu2022}]{schaye1999,rudie2012,bolton2014}. 
Furthermore, the right panel of Fig.~\ref{fig:b_NHI_sim_fit} gives a rough correlation between the $b_\text{fit}$ and $b_\text{thermal}$ and provides an approximate estimation of the strength of the non-thermal broadening of the \lya{} lines at $z=0.1$. It suggests that for the Nyx simulation, the non-thermal contribution to the $b$-parameter can be modelled by a 'turbulent' motion in the IGM with $ b_\text{notherm} \sim 20$ km/s (indicated by the black dash-dot line in Fig.~\ref{fig:b_NHI_sim_fit}).

     \begin{figure*}
 \centering
    \includegraphics[width=1.0\linewidth]{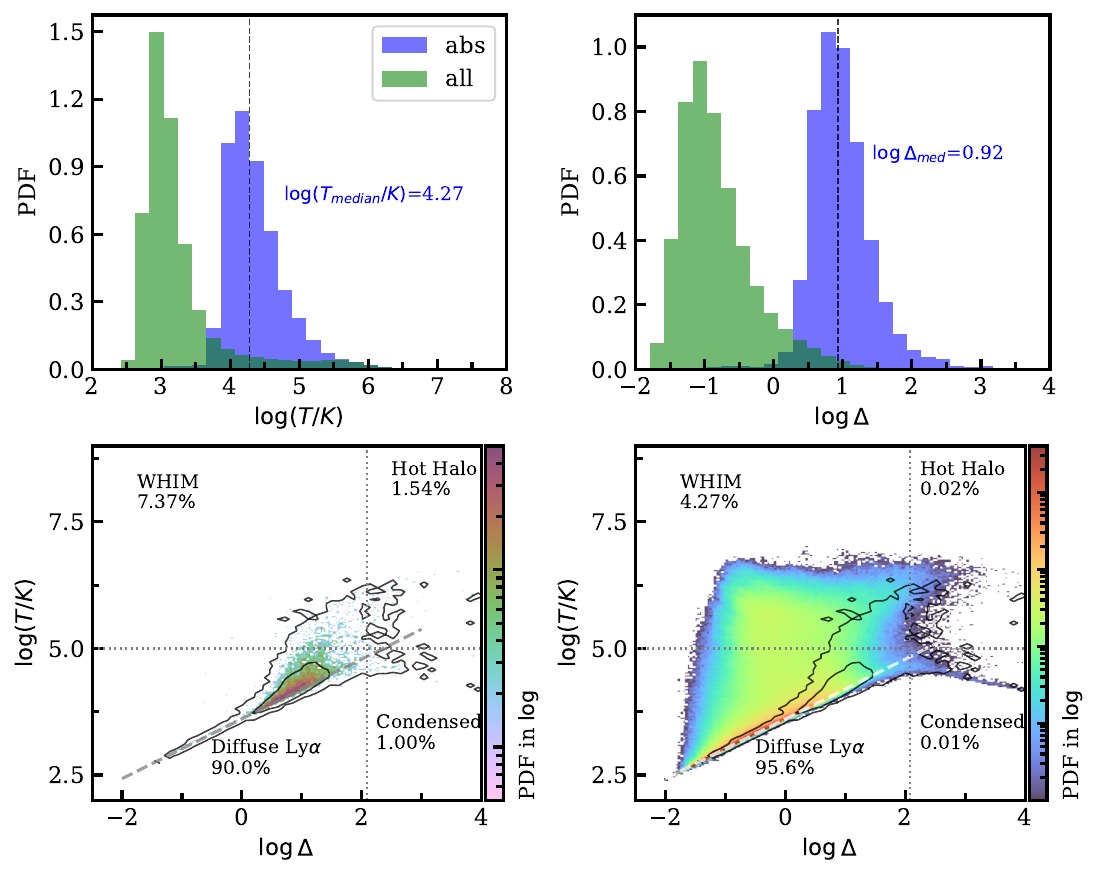}
  \caption{  Distributions of $\Delta$ and $T$ of the simulated \lya{} absorbers in the Nyx simulation, compared with the full simulation.
  The ensemble consists of 10510 absorbers, all obtained from the 1000 spectra discussed in \S~\ref{sec:raw_peaks}.
  The top panels show the 1D distributions of $T$(right) and $\Delta$(left) 
  for the whole simulation (green) compared with simulated \lya{} absorbers (blue). 
  The medians of the $T$ and $\Delta$ for the simulated absorbers are indicated by dashed black lines. 
  The bottom panels plot the 2d $T$-$\Delta$ distributions for the \lya{} absorbers(left) and for the whole simulation(right), while the contours for 1,3-$\sigma$ (68\% and 99.7\%) of the $T$-$\Delta$ distribution of the absorbers are shown in both panels. The volume-weighted gas phases for absorbers and the whole simulation are given in the left panel and the right panel respectively. The best-fit power-law $T$-$\Delta$ relationships are given in the bottom panels as comparisons. }
  \label{fig:raw_peak}
\end{figure*}

We summarize the ($\Delta$, $T$) for the ensemble of simulated \lya{} absorbers identified in the Nyx simulation in Fig.~\ref{fig:raw_peak}.
Considering that we have established one-to-one correspondence between the simulated absorbers and observed (mock) absorption lines, 
we employ a consistent filter to both sets, which selects \lya{} lines with,
$12.5 \leq \log ( N_{\mathHI} / \text{cm}^{-2}) \leq 14.5$ and $0.5 \leq \log ( b / \text{km s}^{-1}) \leq 2.5$ (see \S~\ref{sec:FM_VPFIT}).
In the upper panels, we plot the volume-weighted 1D marginal distributions of $\Delta$ and $T$ for all simulation grid cells, juxtaposed with the 1D distributions of $\Delta$ and $T$ for the simulated \lya{} absorbers,
showing that the simulated \lya{} absorbers, 
in general, have higher temperature and over-density, compared with the full simulation. 
The peaks of the $\Delta$ and $T$ distributions of the simulated \lya{} absorber highlight the specific range of $\Delta$ and $T$ to which the \lyaf{} is sensitive at $z=$ 0.1.
More specifically, the \lyaf{} is most sensitive to the IGM characterized by $\log\Delta = 0.92$ and $T = 10^{4.27}$K. 
It is worth mentioning that, as mentioned in \S~\ref{sec:Grid_label}, the \lya{} optical depth $\tau_{\lya}$ is dependent on $\Gamma_{\mathHI{}}$. Consequently, the regions to which the \lyaf{} is sensitive also depend on $\Gamma_{\mathHI{}}$. This point will be fully discussed later in \S~\ref{sec:GUV_abs}.

The bottom left panel of Fig.~\ref{fig:raw_peak} shows the (volume-weighted) $T$-$\Delta$ distributions for simulated \lya{} absorbers (left), and all grid cells in the simulation (right), while the volume-weighted gas fractions\footnote{As previously mentioned, for each simulated \lya{} absorber, we use the $T$ and $\Delta$ at its $n_{\mathHI}$ peak, which dominates the \lya{} absorption. To this end, when calculating the volume-weighted gas fractions, we do not take the physical size into account, but instead, only consider the one simulation cell where the $n_{\mathHI}$ reaches its maximum. This is reasonable since typical $n_{\mathHI}$ peaks seen in this study are so narrow that most of the $N_{\mathHI}$ comes from the peak cell. As a result, the so-called volume-weighted gas fractions for simulated absorbers are effectively unweighted. This approach is used for all gas fractions related to simulated \lya{} absorbers throughout this paper.} 
are given in annotations for simulated absorbers and the whole simulation in the left and the right panel respectively. The black contours in both panels illustrate the 1 and 3 $\sigma$ (68\% and 99.7\%) contours for the  $T$-$\Delta$ distribution of the simulated \lya{} absorbers.
The $T$-$\Delta$ distribution of the simulated \lya{} absorbers appears to be scattered at low-$z$, extending into the WHIM phase, due to the pervasive effects of shock heating. 
As per the gas phase fractions of the \lya{} absorbers shown in the bottom left panel,  approximately 7\% of the absorbers originate from the WHIM phase, suggesting that the low-$z$ \lyaf{} does probe the WHIM 
(see the right panel of Fig.~\ref{fig:Nyx_lines} as an example), although its sensitivity is notably limited given the small fraction of lines arising from this phase. Such a result aligns with \citet{{Tepper-Garcia2012}} regarding the detectability of the Broad \lya{} Absorbers (BLAs) at low-$z$ under realistic conditions.

 \subsection{ Simulated \lya{} Absorbers in IllustrisTNG and Illustirs}
  \label{sec:abs_ILL}

To further study the effects of the feedback mechanisms on the \lyaf{} at $z=0.1$,
we identify the simulated \lya{} absorbers in both the IllustrisTNG and Illustris simulations, 
and pair them to the corresponding absorption lines present in the mock spectra, following the method outlined in section \ref{sec:raw_peaks}. 
For each simulation,we carry out the identification process across 1000 mock spectra and summarize the physical properties of the simulated absorbers.
It is worth mentioning that here the IllustrisTNG and Illustris simulations are tuned to have identical \dndz{}, which requires 
different $\Gamma_{\mathHI{}}$ values (see \S~\ref{sec:GUV_dNdz}).

       \begin{figure*}
 \centering
    \includegraphics[width=1.00\textwidth]{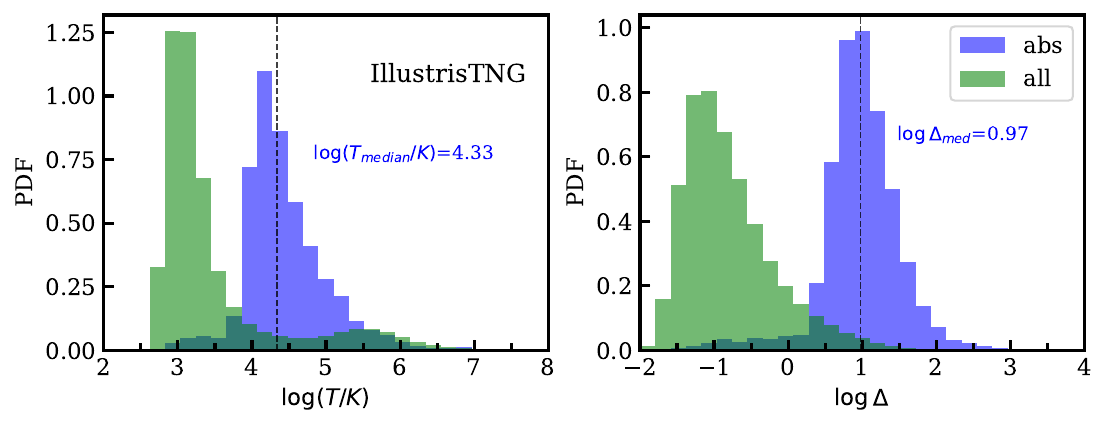}
    \includegraphics[width=1.00\textwidth]{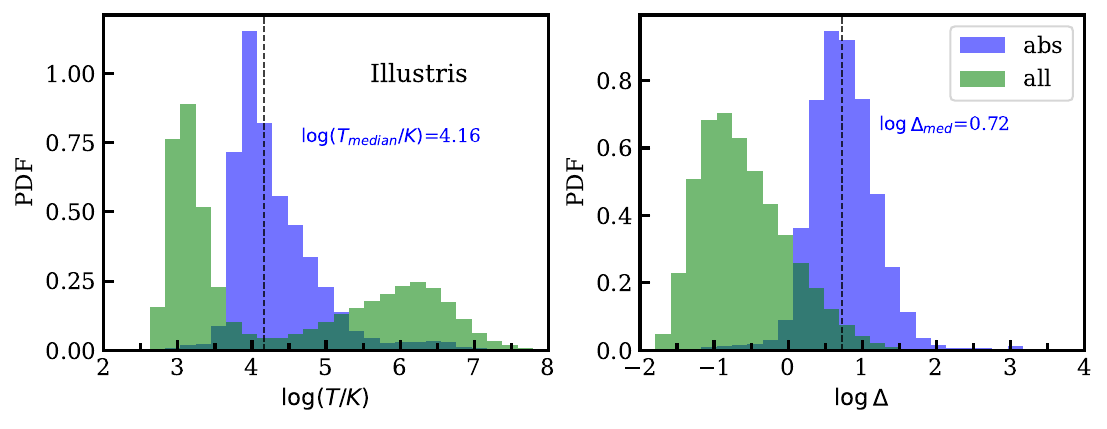}
  \caption{ Marginalized 1D $\Delta$ and $T$ distributions of the simulated \lya{} absorbers in the IllustrisTNG (top) and Illustris (bottom) simulation.
 The medians of the $T$ and $\Delta$ for the simulated absorbers are indicated by dashed black lines. 
  The overall $\Delta$ and $T$ distributions of the full simulations are plotted as comparisons. The two simulations share the same \dndz{}, which is the same value observed in the \citetalias{Danforth2016} dataset.}
  \label{fig:bN_rawpeaks_1dhist}
\end{figure*} 

        \begin{figure*}
 \centering
    \includegraphics[width=1.00\textwidth]{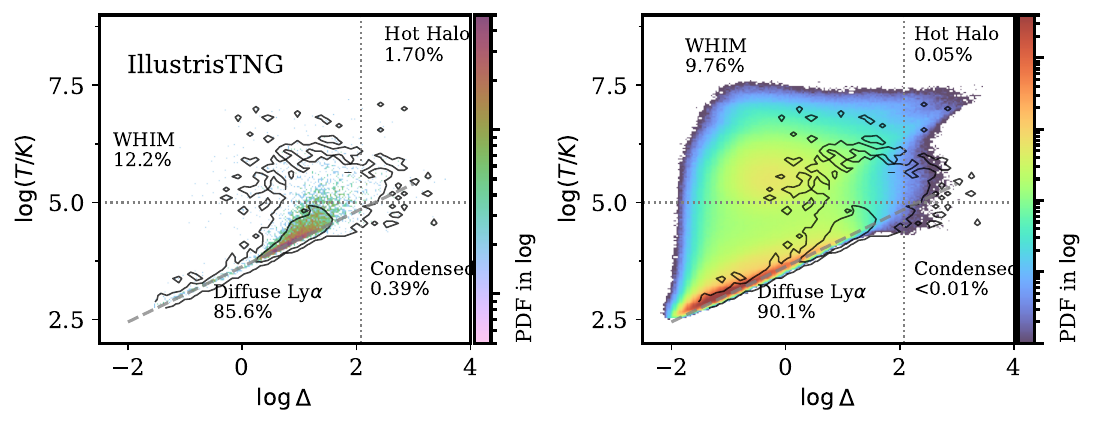}
    \includegraphics[width=1.00\textwidth]{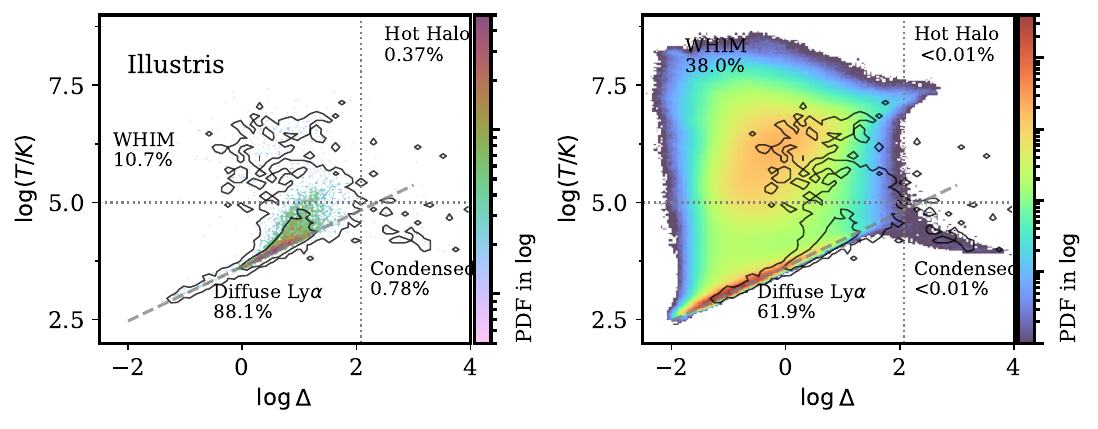}
  \caption{$T$-$\Delta$ distributions of the \lya{} absorbers in the IllustrisTNG (top) and Illustirs (bottom) simulations, compared with the $T$-$\Delta$ distributions of full simulations (right). The contours for 1,3-$\sigma$ (68\% and 99.7\%) of the $T$-$\Delta$ distribution of the absorbers are shown in both panels. The volume-weighted gas phases for absorbers and the whole simulation are given in the left panel and the right panel respectively.  The best-fit power-law $T$-$\Delta$ relationships are given as comparisons. The two simulations are \dndz{} matched.}
  \label{fig:bN_rawpeaks_2dhist}
\end{figure*}

We plot the marginalized 1D distributions of the $\Delta$ and $T$ for both IllustrisTNG (top) and Illustris (bottom) in Fig.~\ref{fig:bN_rawpeaks_1dhist}.
The plots show that the overall distributions of $T$ and $\Delta$ for the two simulations are evidently different due to their different feedback recipes. For instance, the extreme feedback in Illustris simulation results in much more WHIM compared with IllustrisTNG, causing a secondary peak in its $T$ distribution.
However, the distributions of $T$ and $\Delta$ for the absorbers in both simulations are comparable, 
with $\log (T_\text{med}/\text{K}) =$4.33, $\log \Delta_\text{med} =$0.97 for IllustrisTNG, and $\log (T_\text{med}/\text{K}) =$ 4.16, $\log \Delta_\text{med} =$ 0.72 for Illustris.  
Moreover, we discover that the differences in $\log (T_\text{med}/\text{K})$ and $\log \Delta_\text{med}$ for the three simulations are actually caused by the different $\Gamma_{\mathHI{}}$ values used for the three simulations. The relevant discussion is presented in  \S~\ref{sec:GUV_abs}.

The (volume-weighted) 2D $T$-$\Delta$  distributions for simulated \lya{} absorbers in both IllustrisTNG (top) and Illustris (bottom) simulations are shown in the left column of Fig.~\ref{fig:bN_rawpeaks_2dhist}, whereas the (volume-weighted) 2D $T$-$\Delta$ distributions for the whole simulations are given in the right column as comparisons. The volume-weighted gas fractions are given in the annotation for simulated absorbers and full simulation in the left and the right panels respectively.
For the simulated \lya{} absorbers, 12.2\% (10.7\%) of the \lya{} absorbers arise from the WHIM for IllustrisTNG (Illustris), while the value for Nyx simulation is approximately 7\%. The 1 and 3-$\sigma$ (68\% and 99.7\%) contours for the $T$-$\Delta$ distributions for the simulated \lya{} absorbers are also given in the Figure, showing that their $T$-$\Delta$ distributions are more scattered compared with these in Nyx simulation, especially for the WHIM phase absorbers. 
These differences are caused by stronger shock heating in IllustrisTNG and Illustris simulations compared with Nyx simulation, caused by their feedback mechanisms. 
However, while the WHIM fractions for the two simulations are remarkably different, 9.8\% for IllustrisTNG and 38.0\% for Illustris, the WHIM fractions for the \lya{} absorbers are similar, both around 11\%. Furthermore, in \S~\ref{sec:GUV_abs} we demonstrate that the small difference in WHIM fractions for simulated absorbers in the two simulations actually arises from the different $\Gamma_{\mathHI{}}$ values used in the two simulations.
Such a fact implies that the low-$z$ \lyaf{} does not probe the WHIM effectively under realistic conditions, which is consistent with the conclusion drawn by \citet{Khaire2023}.

 \subsection{ Simulations under the same $\Gamma_{\mathHI{}}$}
  \label{sec:GUV_abs}

Considering that the calculation of the \lya{} optical depth $\tau_\text{\lya{}}$ involves $\Gamma_{\mathHI{}}$,
and given that the observed absorption feature (i.e., the \lyaf{}) consistently probes regions with $\tau_{\text{Ly} \alpha} \sim 1$, 
 it follows that the $T$ and $\Delta$ of these regions probed by the \lyaf{}, 
are influenced by the $\Gamma_{\mathHI{}}$ values. 
Such an argument can be qualitatively demonstrated by the fluctuating Gunn-Peterson approximation \citep[FGPA, see][]{Weinberg1997}
\begin{equation}
\tau_{\text{Ly} \alpha} \propto x_{\rm HI} n_{\rm H} \propto \frac{n^2_{\rm  H} T^{-0.7}}{\Gamma_{\rm HI}}  \propto \frac{ {\Delta}^{2.7 -\gamma} }{\Gamma_{\rm HI}}  \propto \frac{ T^{ 2/(\gamma -1) -0.7} }{\Gamma_{\rm HI}},
\label{eq: tau_delta_T}
\end{equation}
where the $\tau_{\text{Ly} \alpha}$ denotes the \lya{} optical depth and the $n_{\rm H}$ is the hydrogen number density.  
Since the \lyaf{} always probes the region with $\tau_{\text{Ly} \alpha} \sim 1$,  the last two terms in eq.~\ref{eq: tau_delta_T} suggest that the $\Gamma_{\mathHI{}}$ is in positive correlation with $\Delta$ and $T$ respectively, given the $\gamma \sim 1.6$ at $z=0.1$.

In our analysis, the three simulations are tuned to match \dndz{}. 
However, due to the degeneracy between $\Gamma_{\mathHI{}}$ and feedback mechanisms, 
each simulation ends up with a distinct $\Gamma_{\mathHI{}}$ value (refer to \S~\ref{sec:GUV_dNdz}). 
As a result, the $T$ and $\Delta$ distributions of the simulated \lya{} absorbers in these simulations are influenced not just by the feedback but also by the varying $\Gamma_{\mathHI{}}$ values.
To isolate and examine solely the impact of feedback, 
we post-process the IllustrisTNG and Illustris simulations to align with the $\Gamma_{\mathHI{}}$ value used in Nyx, set at $ \log (\Gamma_{\mathHI{}}/s^{-1}) = -13.093$. With this consistent $\Gamma_{\mathHI{}}$ across the three simulations, we re-perform the analysis from the prior section and summarize the results below. It is worth mentioning that the overall $T$-$\Delta$ distributions of simulations are determined by the cooling and heating processes during their evolution and are not altered by the post-processing of the $\Gamma_{\mathHI{}}$.

\begin{figure*}
 \centering
    \includegraphics[width=1.0\linewidth]{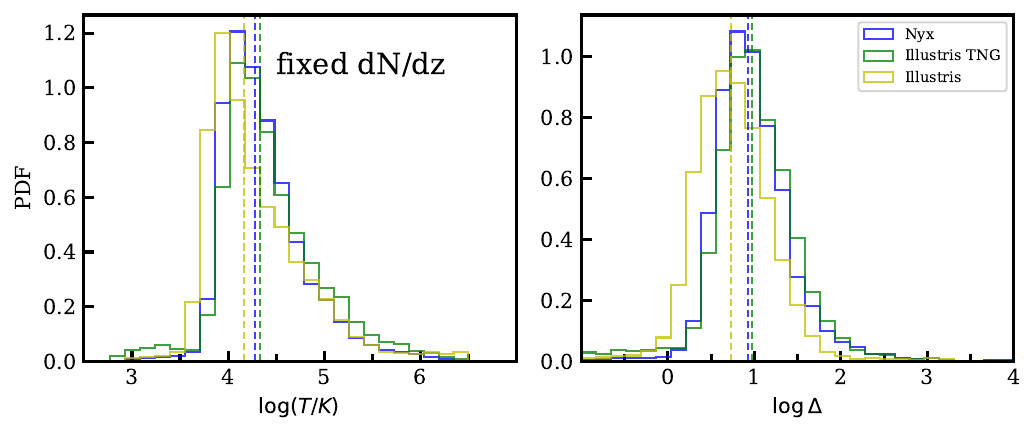}
  \caption{  Marginalized 1d distribution of $T$ (left), and $\Delta$ (right) of the simulated \lya{} absorbers identified in  Nyx (blue), IllustrisTNG (green), and Illustris (yellow) simulations. The medians of $\log T$ and $\log \Delta$ are indicated by vertical dashed lines. The three simulations are turned to have identical \dndz{}.}
  \label{fig:tri_1dhist_tau_dndzmatch}
\end{figure*}

We plot the marginalized $\Delta$ and $T$ distributions and their median values for \lya{} absorbers in Nyx, IllustrisTNG, and Illustris simulations with the same \dndz{} in Fig.~\ref{fig:tri_1dhist_tau_dndzmatch}.
Interestingly, for simulations with the same \dndz{},
the $T$ and $\Delta$ for absorbers are correlated with its  $\Gamma_{\mathHI{}}$. More specifically,   with $\Gamma_{\mathHI{},\text{Illustris}} < \Gamma_{\mathHI{},\text{Nyx}} < \Gamma_{\mathHI{},\text{IllustrisTNG}} $ (see Fig.~\ref{fig:dNdz} ), we obtain $ T_{\text{med,Illustris}} < T_{\text{med,Nyx }} <T_{\text{med,IllustrisTNG}}$ and $ \Delta_{\text{med,Illustris}} < \Delta_{\text{med,Nyx }} <\Delta_{\text{med,IllustrisTNG}}$.

\begin{figure*}
 \centering
    \includegraphics[width=1.0\linewidth]{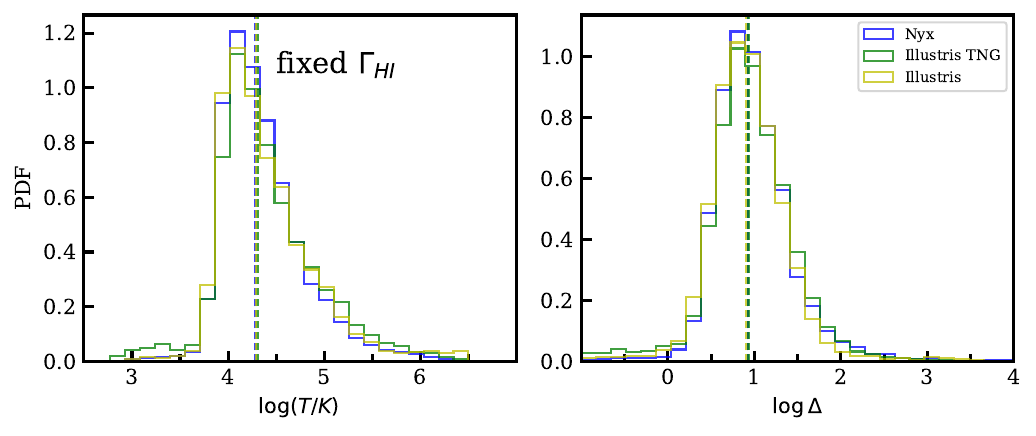}
  \caption{Marginalized 1d distribution of $T$ (left), and $\Delta$ (right) of the simulated \lya{} absorbers identified in  Nyx (blue), IllustrisTNG (green), and Illustris (yellow) simulations. The medians of $\log T$ and $\log \Delta$ are indicated by vertical dashed lines. The three simulations used here are post-processed to have the same UV background photoionization rate, with $\Gamma_{\mathHI{}}$ =-13.093.}
  \label{fig:tri_1dhist_tau}
\end{figure*}

We plot the marginalized $\Delta$ and $T$ distributions and their median values for \lya{} absorbers in the three simulations under the same $\Gamma_{\mathHI{}}$ in Fig.~\ref{fig:tri_1dhist_tau}.
Under the same $\Gamma_{\mathHI{}}$, the $T$ and $\Delta$ distributions for simulated absorbers in all three simulations become almost identical, having nearly the same median values for $T$ and $\Delta$ respectively.  
Such a result suggests that while feedback evidently affects the overall $T$-$\Delta$ distributions of the low-$z$ IGM (see Fig.~\ref{fig:tri_rho_T}), their impacts on the low-$z$ \lyaf{} are not distinguishable under realistic conditions.

In Fig.~\ref{fig:bN_rawpeaks_2dhist}, we plot the 2D $T$-$\Delta$ distributions of the simulated \lya{} absorbers in IllustrisTNG (left) and Illustris (right), under the same $\Gamma_{\mathHI{}}$. 
While the overall IGM $T$-$\Delta$ distributions for the two simulations are evidently different (see Fig.~\ref{fig:tri_rho_T}),
the $T$-$\Delta$ distributions of the simulated \lya{} absorbers in these two simulations are similar, and the gas phase fractions for absorbers in both simulations are almost identical, suggesting that the small difference in the WHIM fractions of the simulated absorbers (see Fig.~\ref{fig:bN_rawpeaks_2dhist}) are caused by different $\Gamma_{\mathHI{}}$ values. Such results indicate that the $\Gamma_{\mathHI}$ has a much stronger impact on the \lyaf{} compared with the feedback mechanisms implemented in IllustrisTNG and Illustris simulations.

        \begin{figure*}
 \centering
\includegraphics[width=0.49\linewidth]{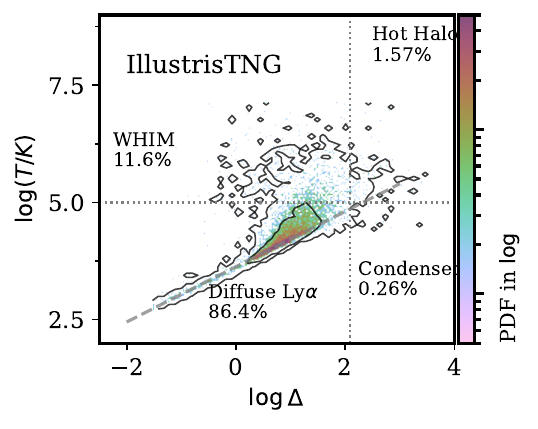}
    \includegraphics[width=0.49\linewidth]{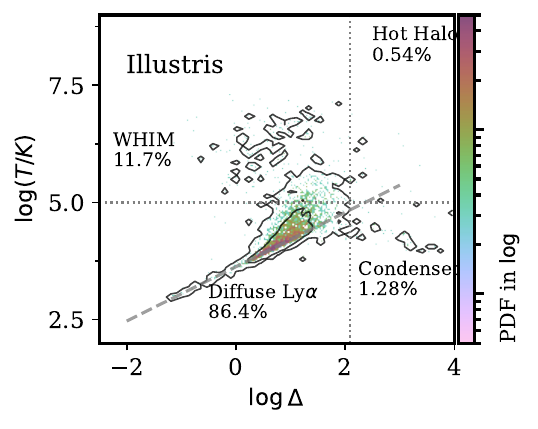}
  \caption{$T$-$\Delta$ distributions of the simulated \lya{} absorbers in the IllustrisTNG (left) and Illustirs (right) simulations under the same $\Gamma_{\mathHI}$. The contours for 1,3-$\sigma$ (68\% and 99.7\%) of the $T$-$\Delta$ distribution of the absorbers are shown in both panels. The volume-weighted gas phase fractions for absorbers in both simulations are given respectively. }
  \label{fig:bN_rawpeaks_2dhist}
\end{figure*}

\section{Summary and Discussion}
  \label{sec:discussion}

In this paper, we explore the effects of the WHIM, which causes significant dispersion in the IGM $T$-$\Delta$ distribution, on the low-$z$ \lyaf{} and the IGM thermal state $[T_0,\gamma]$ measured from it.
We first evaluate the effectiveness of $[T_0,\gamma]$ as IGM parameters under the inference framework presented in \citetalias{Hu2022}, and compare its performance with the photoheating labels [$A$,$B$]. We discover that the thermal state [$T_0$, $\gamma$] still parameterizes the IGM effectively in spite of the dispersion in the IGM $T$-$\Delta$ distribution. 
We further apply the inference method to IllustrisTNG and Illustris simulations which implement different
variants of feedback, potentially making them better approximations to the real Universe. We discover that the [$T_0$,$\gamma$] of these two simulations can be recovered using the inference method within reasonable offsets.
Considering the inference results and the huge difference across the three simulations in the IGM WHIM fractions, we conclude that the \lyaf{} does not probe the WHIM effectively under realistic conditions, and the IGM thermal state [$T_0$,$\gamma$] is not affected by the shock heating caused by AGN feedback and other astrophysical processes significantly at $z=0.1$.
To further confirm our conclusion, we identified the \lya{} absorbers in all three simulations at $z=0.1$, and pair them to the corresponding absorption lines identified in the mock spectra. The physical properties of the simulated \lya{} absorbers support our conclusion that the observable \lyaf{} are not affected by the substantial WHIM in the low-$z$, and the thermal state $[T_0, \gamma]$ measured from the \lyaf{} remains solid.
In this section, we summarize our paper and present our discussion as follows.

\begin{itemize}
 
\item We compare the performance of $[T_0,\gamma]$ as neural network training labels against the photoheating labels [$A$,$B$], i.e. the photoheating rate rescaling factors used to generate the Nyx simulation suite with various thermal histories. Given that the $[A,B]$ parameters were actually used to generate the simulation outputs, one might expect that they would serve as a better set of labels than $[T_0,\gamma]$. 
However, the inference results show the efficacy of these two sets of labels are comparable, suggesting that the [$T_0$, $\gamma$], which parameterize the power law $T$-$\Delta$ relationship, still effectively characterize the \lya{} observables at low-$z$, notwithstanding the dispersion in the $T$-$\Delta$ distribution induced by shock heating at low-$z$. 

\item We explored the degree to which the presence of feedback can influence or bias the inference of the 
IGM thermal state parameters from the  \bndist{}. In the context of our inference framework, this question becomes: what would happen if we used a simulation grid without feedback to infer the thermal state of a  Universe that has strong feedback? Would the feedback lead to biased inference?
To address these questions, we apply our inference procedure trained on Nyx simulations without feedback to
mock datasets from the IllustrisTNG and Illustris simulations which include feedback, whereby the latter
serve as potential proxies for the real Universe.  We find that the [$T_0$,$\gamma$] of IllustrisTNG and Illustris can be recovered within small offset, where  $\Delta \log(T_0/\text{K}) = -0.022$ dex, $\Delta \gamma = 0.064$ for IllustrisTNG and $\Delta \log(T_0/\text{K}) = 0.047$ dex, $\Delta \gamma = -0.094$ for Illustris.
These offsets are smaller than the typical precision afforded by a realistic dataset, i.e., $ \Delta \log T_0 \lesssim 0.5$ $\sigma_{\log T_0}$, and $\Delta \gamma \lesssim \sigma_ {\gamma}$.

\item We developed a method to identify regions in the simulation responsible for the \lya{} absorption
lines identified via Voigt-profile fitting, allowing us to determine their temperature $T$ and overdensity $\Delta$
from the simulation skewers. 
For the Nyx simulations, the simulated \lya{} absorbers have a median density $\log \Delta_\text{median} = 0.92$, a median temperature $T_\text{median} = 10^{4.27}$K, and about 7\% of the simulated \lya{} absorbers
have $T > 10^5$, making them outliers from 
the power-law $T$-$\Delta$ relationship. This low fraction is consistent with the previous study of \citet{Tepper-Garcia2012} on the low-$z$ BLAs.

\item As pointed out in previous work \citep{Bolton22,Khaire2023,Tillman2023b}, the \lyaf{} is affected by the UV background, which impacts the \dndz{}. Nevertheless, we observe that the temperature and overdensity of the region probed by the low-$z$ \lyaf{} are also affected by the UV background photoionization rate $\Gamma_{\mathHI{}}$ used in the simulation. 
For \dndz{} matched simulations, the $T$ and $\Delta$ of the simulated \lya{} absorbers are correlated with its $\Gamma_{\mathHI{}}$ respectively. Specifically, the \lyaf{} probes regions with higher $T$ and $\Delta$ given a higher $\Gamma_{\mathHI{}}$. 
This is because for \lya{} absorbers with $\tau_{\lya{}} \sim$1, the fluctuating Gunn-Peterson approximation implies that $\Gamma_{\mathHI} \propto {\Delta}^{2.7 -\gamma} \propto  T^{ 2/(\gamma -1) -0.7}$, where $\gamma \sim 1.6$.

\item We post-processing the three simulations to share the same $\Gamma_{\mathHI{}}$, allowing us to explore the effects of different mechanisms.  Under the same $\Gamma_{\mathHI{}}$, the $T$ and $\Delta$ distributions of the simulated \lya{} absorbers across all three simulations become almost indistinguishable, converging to nearly identical median values, while the overall IGM $T$-$\Delta$ distributions remain different among the simulations, due to their distinct feedback mechanisms. 
For the WHIM fractions, the volume-weighted WHIM fractions for IllustrisTNG and Illustris stand at 9.8\% and 38.0\%, respectively, but the WHIM fractions for the simulated \lya{} absorbers in both simulations are nearly identical, averaging around 11.6\%.
This suggests that while feedback significantly alters the low-$z$ IGM $T$-$\Delta$ distribution, especially the WHIM phase gas, their impacts on the low-$z$ \lyaf{} is indistinguishable under realistic conditions. Such a conclusion aligns with the results derived from various statistics of the low-$z$ \lya{} forest by \citet{Khaire2023}.

We have thus far demonstrated the robustness of the thermal state $[T_0,\gamma]$ as IGM parameters at low-$z$, in spite of the dispersion in the $T$-$\Delta$ distribution induced by shock heating. We also proved that the \citetalias{Hu2022} inference framework can effectively measure the thermal state  $[T_0,\gamma]$ notwithstanding the feedback mechanisms implemented in the IllustrisTNG and Illustris simulations.  
Looking ahead, we plan to apply the \citetalias{Hu2022} inference methodology to simulations with more flexible and sophisticated feedback mechanisms, including EAGLE \citep{EAGLE} and CAMELS suite \citep{camels_presentation}.
The outcomes will provide us with a deeper understanding of the impact of various feedback processes on low-$z$ IGM.
Moreover, by applying our methodology on archival \ac{HST} \ac{COS} and \ac{STIS} datasets, we expect precise measurements of the low-$z$ \ac{IGM} thermal state.
These results will pinpoint the onset of the discrepancy in the $b$ parameters of the low-$z$ \lyaf{} between current simulations and observations, which is essential for unravelling the underlying physics and acquiring a comprehensive picture of the IGM thermal evolution at low-$z$ after the epoch of helium reionization.

 \end{itemize}


\begin{acronym}
	\acro{AGN}{active galactic nuclei}
	\acro{CMB}{Cosmic Microwave Background}
	\acro{COS}{Cosmic Origins Spectrograph}
	\acro{DELFI}{density-estimation likelihood-free inference}
	\acro{DM}{dark matter}
	\acro{DLA}{damped Ly$\alpha$}
	\acro{GP}{Gaussian process}
	\acro{HIRES}{High Resolution Echelle Spectrometer}
	\acro{HST}{Hubble Space Telescope}
	\acro{IGM}{intergalactic medium}
	\acro{KDE}{Kernel Density Estimation}
	\acro{KODIAQ}{Keck Observatory Database of Ionized Absorbers toward QSOs}
	\acro{LD}{least absolute deviation}
	\acro{LLS}{Lyman limit systems}
	\acro{LS}{least squares}
	\acro{LSF}{line spread function}
	\acro{MCMC}{Markov chain Monte Carlo}
	\acro{MW}{Milky Way}
	\acro{NDE}{neural density estimation}
	\acro{PCA}{principal component analysis}
	\acro{PDF}{probability density function}
	\acro{PKP}{\ac{PCA} decomposition of \ac{KDE} estimates of a \ac{PDF}}
	\acro{QSO}{quasi-stellar objects}
	\acro{SNR}{signal-to-noise ratio}
	\acro{STIS}{space telescope imaging spectrograph}
	\acro{TDR}{temperature-density relation}
	\acro{THERMAL}{Thermal History and Evolution in Reionization Models of Absorption Lines}
	\acro{UV}{ultraviolet}
	\acro{UVB}{ultraviolet background}
	\acro{UVES}{Ultraviolet and Visual Echelle Spectrograph}
	\acro{WHIM}{warm hot intergalactic medium}
\end{acronym}

\section*{Acknowledgements}

We thank the members of the ENIGMA\footnote{http://enigma.physics.ucsb.edu/}, 
Siang Peng Oh, Timothy Brandt, and K.G. Lee for helpful discussions and 
suggestions.

Calculations presented in this paper used the hydra and draco clusters
of the Max Planck Computing and Data Facility (MPCDF, formerly known
as RZG). MPCDF is a competence center of the Max Planck Society
located in Garching (Germany).
This research also used resources of the National Energy Research Scientific Computing Center (NERSC), a U.S. Department of Energy Office of Science User Facility located at Lawrence Berkeley National Laboratory, operated under Contract No. DE-AC02-05CH11231 
In addition, we acknowledge PRACE for awarding us access to JUWELS hosted by JSC, Germany.
JO acknowledges support from grant CNS2022-135878 from the Spanish Ministerio de Ciencia y Tecnologia.

\section*{Data Availability}

The simulation data and analysis code underlying this article will be shared on reasonable request to the corresponding author.




\bibliographystyle{mnras}

\bibliography{references.bib}


\appendix

\section{ Inference based on the Photoheating labels [$A$,$B$]}
\label{sec:label_AB}

In this section, we present our inference results using the framework where different Nyx models are labelled by the photoheating parameters [$A$,$B$] instead of the thermal state $[T_0,\gamma]$, and the inference method returns $[A,B,\log \Gamma_{\mathHI}]$.
The inference is conducted following the procedures described in \S~\ref{sec:inference}, based on the DELFI \bndist{} emulator trained on training dataset labelled by [$A$,$B$, $\log \Gamma_{\mathHI{}}$], which returns $P ( b \mathbin{,} N_{\mathHI{}} \, \given \, A, B, \log \Gamma_{\mathHI{}})$.

\begin{figure}
 \centering
    \includegraphics[width=\columnwidth]{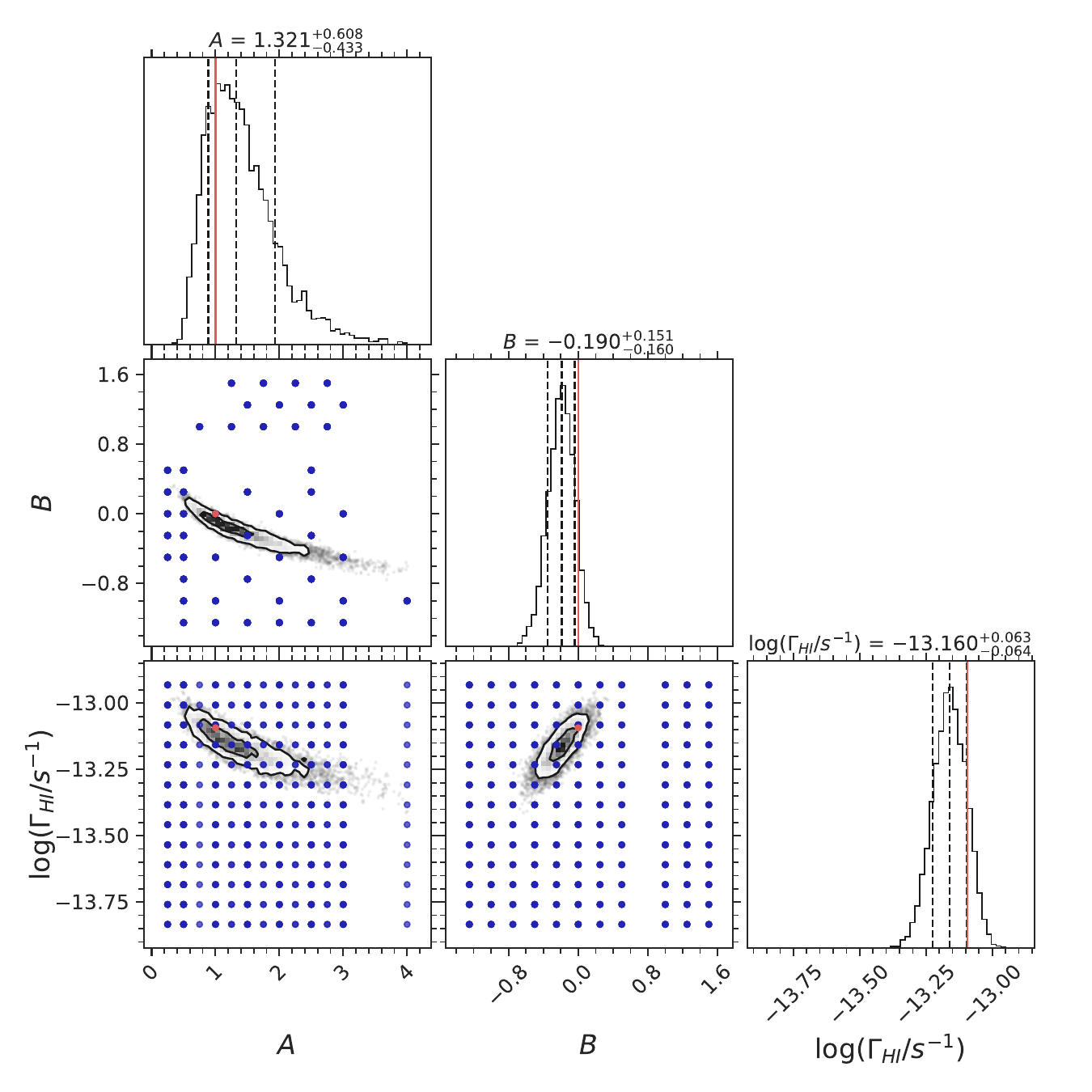}
  \caption{An example of posterior obtained by our inference method based on [$T_0$, $\gamma$, $\Gamma_{\mathHI()}$]. Projections of the thermal grid used for generating models are shown as blue dots, while the true model is shown as red dots. The inner (outer) black contour represents the projected 2D 1(2)-sigma interval. The parameters of true models are indicated by red lines in the marginal distributions, while the dashed black lines indicate the 16, 50, and 84 percentile values of the posterior. The true parameters are: {$A$} = 1.0 and $B =$0.0, {$\log (\Gamma_{\mathHI{}} /\text{s}^{-1})$} = -13.093. }
  \label{fig:Nyx_corner_AB}
\end{figure}

\begin{figure}
 \centering
    \includegraphics[width=\columnwidth]{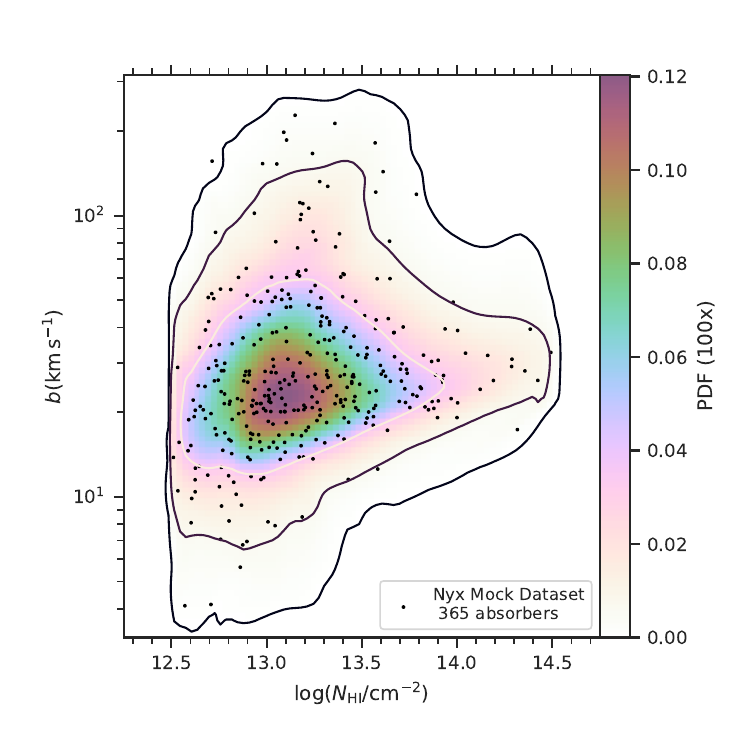}
  \caption{The colour map is the full \bndist{} recovered from the Nyx mock dataset, which is emulated by our DELFI emulator based on the best-fit parameters (median values of the marginalized MCMC posterior), where $A$ = 3.695 (1.0) and $B=1.507$ (0.0) and {$\log (\Gamma_{\mathHI{}} /\text{s}^{-1})$} =-13.237 (-13.093), the true parameters are given in parentheses. Black dots are the mock datasets we used in the inference. 
  The contours correspond to cumulative probabilities of 68\%, 95\% and 99.7\%.  
  For illustration purposes, the values of pdf are multiplied by 100 in the colour bar.}
  \label{fig:fit_Nyx_AB}
\end{figure}

\begin{figure}
 \centering
     \includegraphics[width=1.0\columnwidth]{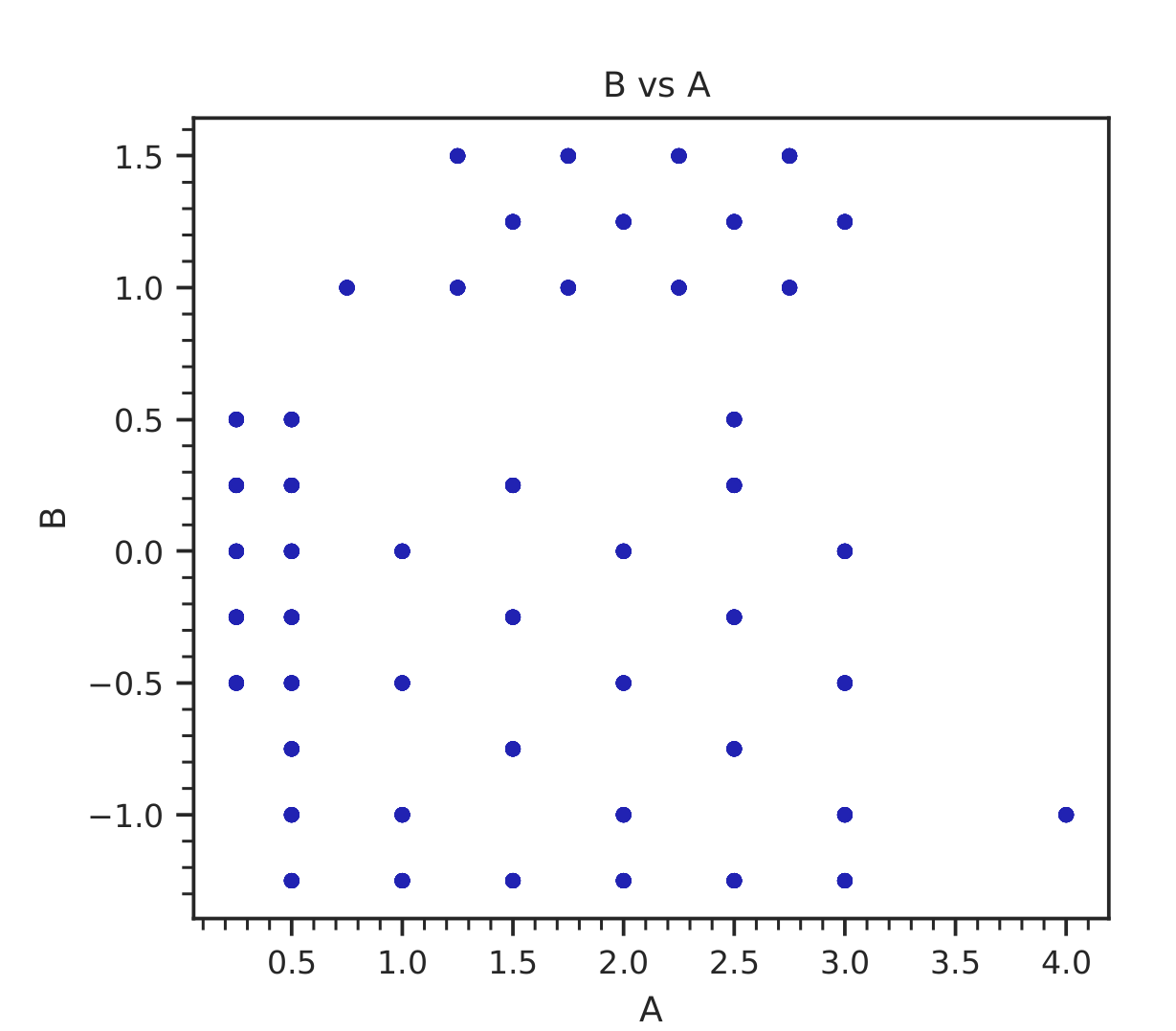}
  \caption{Parameters grid (blue circles) from snapshots of hydrodynamic simulations of the \ac{THERMAL} suite at $z=0.1$ parameterized by the thermal state [$A$,$B$]}
  \label{fig:z01_AB_grid}
\end{figure}

The simulation grid, parameterized by the photoheating labels [$A$,$B$], is given in Fig.~\ref{fig:z01_AB_grid}.
An example of the MCMC posterior obtained based on the aforementioned likelihood function is given in Fig.~\ref{fig:Nyx_corner_AB}. 
The inference method returns $A=$1.321 (1.0), $B=-0.190$ (0.0), $\Gamma_{\mathHI}=-13.160$ (-13.093), whereas the true values are given in the parentheses.
The posterior appears compact, with the medians of the marginalized posteriors landing within 1-$\sigma$ errors for all three parameters.
The \bndist{} recovered from the mock dataset is presented in Fig.~\ref{fig:fit_Nyx_AB},
which is emulated by our DELFI \bndist{} emulator based on the inferred parameters.

We perform an inference test following the \S~\ref{sec:inf_test}, in which we also exclude models that are too close to the parameter boundaries to avoid the truncation of the resulting posteriors.  Specifically, we only use models with $ 3.3 < \log (T_0/\text{K}) < 3.9$, $ 1.0 < \gamma <2.3$, $ -13.75 < \log (\Gamma_{\mathHI{}} /\text{s}^{-1})<-13.0$.
The result of the inference test is shown in Fig.~\ref{fig: inference_test}. The performance looks comparable to the one based on the thermal state $[T_0,\gamma]$, suggesting that $[T_0,\gamma]$ are still effective IGM parameters at low-$z$, notwithstanding the substantial dispersion in the IGM $T$-$\Delta$ distribution induced by pervasive shock heating at this redshift.

\bsp	
\label{lastpage}
\end{document}